\documentclass[APS,prd,amssymb,amsmath,eqsecnum,showpacs,twocolumn] {revtex4}

\usepackage{hyperref}
\usepackage{color}
\usepackage{footmisc} 
\usepackage{array}
\usepackage[usenames,dvipsnames,svgnames,table]{xcolor}
\usepackage{soul,ulem}
\usepackage{enumerate}

\newcommand{\mc}{\mathcal}
\newcommand{\f}[2]{\frac{#1}{#2}}

\newcommand{\kron}[2]{\delta^{#1}_{\phantom{#1}#2}}

\newcommand{\chr}[2]{\Gamma^{#1}_{#2}}
\newcommand{\p}{\phantom}

\begin{document}

\title{Affine-null metric formulation of General Relativity  {at two intersecting null hypersurfaces}}

\author{Thomas M\"adler${}^{1}$
       }
\affiliation{
${}^{1}$  N\'ucleo de Astronom\'ia, Facultad de Ingenier\'ia y Ciencias, Universidad Diego Portales, Av. Ej\'ercito 441, Santiago, Chile \\
\today 
	 }
\email{thomas.maedler@mail.upd.cl}

\begin{abstract}
We revisit Winicour's affine-null metric initial value formulation of General Relativity, where the  characteristic initial value formulation is set up with a  null metric having two affine parameters. In comparison to past work, where the application of the formulation was aimed for the  {timelike-null} initial value problem, we consider here a boundary surface that is a null hypersurface. 
All of the  initial data are either metric functions or first derivatives of the metric. 
Given such a set of initial data, Einstein equations can  be integrated in a hierarchical manner, where first a  set of equations is solved hierarchically on the null hypersurface serving as a boundary. 
Second, with the obtained boundary values, a set of  differential equations, similar to the equations of the Bondi-Sachs formalism, comprising of hypersurface and evolution equations is solved  hierarchically to find the entire space-time metric. 
An example is shown how the double null Israel black hole solution arises after specification to spherical symmetry and vacuum. 
This black hole solution is then discussed to with respect to Penrose conformal compactification of spacetime.
\end{abstract}

\maketitle
\newpage
\section{Introduction}
The characteristic initial value problem of General Relativity  may be expressed in various different formulations and it has brought to light many different aspects and properties of the theory of General Relativity. 
A selective list of examples for such breakthroughs are the Bondi-Sachs mass loss formula \cite{Bondi,Sachs}, the discovery of the asymptotic Bondi-Metzner-Sachs group (BMS) \cite{Bondi,Sachs,SachsBMS,NPbms}, first long-term stable numerical  evolutions of black holes \cite{news,LongStable1,LongStable2}, critical collapse \cite{critical1,critical2}, horizon formation of super-symmetric Yang Mills fields \cite{CheslerYaffe} as well as  mathematical proofs on existence and uniqueness of solutions of Einstein equations \cite{Friedrich_1982,Rendall,ChoquetBruhat}.
The variables used to formulate the characteristic initial value problem may be the metric  { \cite{Bondi,Sachs,Sachs_civp,newt}}, a conformal metric \cite{tam}, null tetrads \cite{NP,GPH,FS_NP}, spinors \cite{Friedrich_1982} or extrinsic curvatures \cite{DInverno,Brady}. 
All of these formulations have in common that there is one family of null hypersurfaces filling the domain of spacetime to be considered. 
If $\mc{N}_{x^0}$ is such a family of null hypersurfaces for which the scalar $x^0$ is constant along each null hypersurface of the family,  and furthermore $x^A=(x^2,x^3)$ are two additional coordinate scalars  chosen to be constant along the generators of $\mc{N}_{x^0}$, then the most general metric at such a family of null hypersurfaces is 
\begin{eqnarray}
\label{eq:gen_null}
g_{ab}dx^a dx^b &=& [g_{00}dx^0 + 2g_{01}dx^1 + 2g_{0A} dx^A]dx^0\nonumber\\
&&+g_{AB}dx^Adx^B\;\;.
\end{eqnarray}
Due to this parameterisation only $x^1$ varies along the generators of the hypersurfaces $x^0=constant$.
Writing out the twice contracted Bianchi identities for \eqref{eq:gen_null} and specifying some of the field equations as main equations that are assumed to hold on the family $\mc{N}_{x^0}$, one finds the so-called Bondi-Sachs lemma \cite{Bondi,Sachs,Sachs_civp,tam} (and also App. \eqref{app:gen_null}). 
The crucial message from this lemma is that there is a set of field equations, the so-called supplementary equations, which hold everywhere on $\mc{N}_{x^0}$ provided they hold for one coordinate value of $x^1$, say $x^1=0$ w.l.o.g.,  on each of the null hypersurfaces of $\mc{N}_{x^0}$. 
The supplementary equations are the set of field equations of \eqref{eq:gen_null} that need to be discussed for the (3-dimensional) boundary surface $\mc{B}$ with $x^1=0$.

Regarding numerical investigations employing a metric of type \eqref{eq:gen_null} there are in principle four different versions to set up an initial-boundary value algorithm: 
\begin{itemize}
  \item [(i)] the timelike-null formulation, 
  \item [(ii)] the vertex-null formulation, 
  \item [(iii)] the double-null formulation or 2+2 formulation, 
  \item [(iv)] affine-null formulation.
\end{itemize}
See, e.g \cite{JeffLRR}, where different numerical realisations are discussed. 
In the timelike-null formulation, the boundary $\mc{B}$ is a world tube of finite size and the family $\mc{N}_{x^0}$ is attached to the exterior of this world tube. 
It is employed in the PITT code\cite{PITT} using a conformally compactified \cite{tam} Bondi-Sachs metric\cite{BSscolar}.
The computational infrastructure is used to solve Einstein equations along outgoing null hypersurfaces. 
The boundary data on $\mc{B}$ are provided by a Cauchy evolution, i.e. an evolution scheme solving Einstein equations with a $3+1$ formalism \cite{BaumgarteShapiro}, in the interior of the world tube. 
The vertex-null formulation is in fact a specialisation of (i), where the world tube collapses to a single world line. 
As a result, the null hypersurfaces become outgoing null cones with vertices on the  {world line}. 
Here the data to `evolve' the families of light cones along the word line must be provided by regularity conditions along the world line \cite{ChoquetBruhat,TMvertex}.
The vertex-null formulation needs to be used if a full characteristic formulation of General Relativity is used for studying compact material sources, e.g. a single star like in \cite{Papadopoulos,Siebel}.
In  double null foliations, $\mc{B}$ is a null hypersurface, on which the data are prescribed.
In particular, the entire spacetime is foliated with respect to pairs of intersecting null hypersurfaces \cite{Hayward,DInverno,Brady}. 
As the intersection of two 3-dimensional null hypersurfaces is a (spacelike) 2-dimensional subspace, the double null formulation is  often referred to as 2+2 foliation.
This constrains \eqref{eq:gen_null} such that there is $g_{00}=0$ everywhere, in addition to $g_{11}=0$ everywhere.
The affine null foliation is characterised by $|g_{01}|=1$, so that both coordinates $x^0$ and $x^1$ are affine parameters \cite{Win_affine}. 
It is  hybrid  with respect to setting up a characteristic-boundary value problem, in the sense that it  can be used for both a time-like boundary or a null boundary surface. 

In this article, we revisit the  affine null-metric formulation. 
In his seminal article  {\cite{Win_affine}}, Winicour shows how the hypersurface equations on outgoing null hypersurfaces can be cast into a hierarchical system of differential equations along the rays generating these null hypersurfaces. 
Having the application of Cauchy-Characteristic Extraction in mind, $\mc{B}$ is a worldtube of finite size. 
In \cite{Win_affine} it is left open how the boundary equations look like, the author instead points to the relevant article  \cite{worldtube}, where a formalism for an evolution scheme along the timelike boundary data is presented. 
If $\mc{B}$ is horizon, equations on the null boundary $\mc{B}$ are presented in \cite{CompleteNull}. 
Regarding main equations exterior to $\mc{B}$, \cite{CompleteNull} did not use the  Einstein equations for an affine null metric but those arising from a Bondi-Sachs metric. 
Here, we formulate the characteristic initial-boundary value problem for the affine, null-metric formulation, where the boundary data are supplied on a null hypersurface.
We find that the free initial data in this formulation consist of 
(i) three scalar functions on the common intersection $\Sigma$ of  the null boundary $\mc{B}$ with an initial null hypersurface  {$\mc{N}_{0}:=\mc{N}_{x^0=0}$}, 
(ii) one 2-vector field on $\Sigma$, 
(iii) one transverse-traceless 2-tensor field on $\Sigma$, 
(iv) one transverse-traceless 2-tensor field on $\mc{B}$ and $\Sigma$ 
and (v) one  transverse-traceless 2-tensor field on $\mc{N}_{0}$ and $\Sigma$.
The two 2-tensor fields are the shear of the null hypersurfaces $\mc{B}$ and $\mc{N}_{0}$, respectively. 
The data on $\Sigma$ determine a further scalar function, a mixed second derivative, $\mu$, of a conformal factor of the 2-metric $g_{AB}$, via an algebraic relation. 
From the eleven degrees of freedom in the initial data, eight and $\mu$ are propagated along the boundary $\mc{B}$ via a hierarchical set of differential equations along the rays forming $\mc{B}$.    
These nine boundary fields on $\mc{B}$ and combinations thereof provide the start values for nine hierarchically-ordered hypersurface equations on the null hypersurfaces $x^0=const$ that determine five of the six physical degrees of freedom of \eqref{eq:gen_null} for values values $x^1>0$.
The remaining missing metric field, which is $g^{11}$,  is subsequently determined algebraically from the hypersurface variables.
Due to the fact that  $x^0$ is affine parameter on $\mc{B}$ and $x^1$ is an affine parameter everywhere along the rays of $\mc{N}_{x^0}$, the gauge condition $g^{11}_{\p{11},1}|_{\mc{B}}=0$ must be assured on each of the null hypersurfaces of the family $\mc{N}_{x^0}$ after each integration  of the hypersurface hierarchy.
This gauge condition eliminates some of the degrees of freedom in the initial data by algebraic combination thereof.   
Once the  data on an initial data surface   {$\mc{N}_{0}$} are determined, the transverse-traceless part of the intrinsic metric of $\mc{N}_0$  is evolved from $\mc{N}_0$ to $\mc{N}_{x^0>0}$ by a simple first order differential equation whose source term depends on the solution of the hypersurface equations. 
This is in difference to the characteristic evolution in the Bondi-Sachs formalism \cite{BSscolar}, where the shear of $\mc{N}_{x^0}$ is propagated between null hypersurfaces for different values of $x^0$. 

In Sec.~\ref{sec:em}, we discuss the electromagnetic analogue for the affine null-metric formulation at two null hypersurfaces. 
This simple case already incorporates most of the important features of the corresponding affine metric formulation. 
The metric for affine null metric formulation is introduced in Sec.~\ref{sec:dn_system}, where we also  discuss the coordinate transformations on $\mc{B}$ leaving the metric on the boundary  unaltered.
Sec.~\ref{sec:EE} presents the main and supplementary equations as they follow from an affine null metric.
The hierarchical sets of differential equation on $\mc{B}$ and $\mc{N}_{x^0}$ as well as the evolution equations are derived in Sec.~\ref{sec:hierarchy}. 
In the subsequent  section Sec.~\ref{sec:maxsym}, the new set of equation is specified and solved for the case where the metric has either spherical or hyperboloidal symmetry.
As solutions, we will determine the metric of flat space in a double null foliation and a generalisation of the double null Israel black hole solution. 
For the particular choice of spherical symmetry, we will then discuss different strategies for the Penrose conformal compactification of the Israel black hole.
Appendix App.~\ref{sec_genRic} contains an abridged presentation of the  {(rather tedious)} calculation of the relevant components of the Ricci tensor for the main  and supplementary equations  of  the metric \eqref{eq:gen_null}.
This general set of equations is then specified for the conformally decomposition $g_{AB} = r^2h_{AB}$ of the 2-metric $g_{AB}$ in appendix App.~\ref{sec_confRic}. 
The obtained Ricci tensor components had been previously used (with the corresponding specification of the metric) for the main equations of the Bondi-Sachs formalism in \cite{BSscolar}.
Here the same Ricci tensor components yield the respective components for  the vacuum field equations in the affine, null metric formulation after specification of the metric to an affine null metric.

We use the MTW conventions \cite{MTW}  {for symmetrisation/antisymmetrisation}, for the curvature and its related  quantities and geometrised units, $c=G=1$.

\section{An electromagnetic example}\label{sec:em}
Features of a  null formulation may be seen from analysing the electromagnetic field in a Minkowski vacuum. 
Taking the standard Minkowski metric  $\eta_{ab} = \mathrm{diag}(-1,1,1,1)$ in Cartesian coordinates $y^a = (t,y^i) {,\; i\in \{1,2,3\} }$, the double null version of the Minkowski metric is found via the coordinate transformation to retarded time $u = t-r$, advanced time $v=t+r$ with $r^2 =\delta_{ij} y^iy^j$ and two angular coordinates $x^A {=(\theta,\phi)}$ parameterising the three dimensional unit vector $r_i(x^A):=r_{,i}$. 
In coordinates $(u,v, x^A)$,  the line element of $\eta_{ab}$ takes the form 
\begin{equation}
\label{metric_dn_mink}
\eta_{ab}dx^adx^b = -dudv+ r^2(u,v)q_{AB}(x^C) dx^Adx^B\;\;,\;\;
\end{equation}
with $r =  \f{1}{2}(v-u)$, $\det(q_{AB}) = q(x^A)$ and the unit sphere metric $q_{AB} (x^C)= \delta_{ij}r^i_{,A}r^j_{,B}$. 
The affine (double) null character of this metric is seen by the absence of the terms $\eta_{uu}$ and $\eta_{vv}$. 
The surfaces $du=const$ and $dv=0$ are both 3-dimensional null hypersurfaces, that is, their normal vector{s are null vectors} and self-orthogonal. 
The intersection of a hypersurface $du=const$ and a hypersurface $dv=const$ is a spherical cross section. 
Let $V_0$ be the ingoing null hypersurface for which $v=v_0$ and $U_0$ be the outgoing null hypersurface for which $u=u_0$.
The common intersection of $V_0$ and $U_0$ is called $S_0$. 
In particular, functions 
\begin{eqnarray*}
\mbox{on } &S_0&\mbox{ depend on the angles}\;\; x^A\;\;,\\
\mbox{on } &V_0&\mbox{ depend on three coordinates}\;\; (u, x^A)\;\; \mbox{and} \\
\mbox{on } &U_0&\mbox{ depend on three coordinates}\;\; (v,x^A)\;\;.
\end{eqnarray*} 
Let  {$F_{ab}=2\nabla_{[a}A_{b]}$} be the Faraday tensor and $A_a$ the four potential of the electromagnetic field. 
The vector field $A_a$ has  the gauge freedom $A_a\rightarrow A_a+\partial_a\chi$, so that we can choose $A_v = 0$ everywhere by the gauge transformation $\chi = -\int_{v_0}^v A_v dv^\prime$.
As remaining  gauge freedom of $\chi$ we may choose  $A_u|_{v=0}=0$. 
Therefore, $A_u = O(v)$ at $v=0$ and consequently $\partial_A A_u|_{v=0} = 0$, but  $\partial_v  A_u|_{v=0} \neq 0$. 
Defining $M^a := \nabla_bF^{ab}$, Maxwell equations in vacuum are given by $M^a = 0$ and the asymmetry of $F_{ab}$ implies $\nabla_a M^a = 0$ so that
\begin{equation}
\label{em_bianchi}
0=  \f{1}{r^2}(r^2M^u)_{,u}+\f{1}{r^2}(r^2M^v)_{,v}+\eth_A M^A
\end{equation}
where $\eth_A$ is the covariant derivative with respect to the unit sphere metric $q_{AB}$.
Designating  $M^u=0$  and $M^A$ as main equations 
which are assumed to hold everywhere,
the conservation condition \eqref{em_bianchi} implies
\begin{equation}
\label{ }
0=\f{1}{r^2}(r^2M^v)_{,v}\;\;.
\end{equation}
So that provided $r\ne 0$, we find $M^v=0$ holds for all values of $v$ provided $M^v=0$ on hypersurface $v=v_0$.
Denoting $\eth^A = q^{AB}\eth_B$, the relevant Maxwell equation, $M^v=0$, is 
\begin{equation}
\label{eq:suppl_em}
 (r^2  A_{u,v})_{,u}\Big|_{v=v_0} 
=  \f{1}{2} \eth^B( A_{B,u})\Big|_{v=v_0}  
\end{equation}
while the $M^u=0$ equation is
\begin{equation}
\label{eq:hyp_em}
(r^2 A_{u,v})_{,v} = \f{1}{2}  \eth^B (A_{B,v})\;\;\,
\end{equation}
and the $M^A=0$ equation is
\begin{eqnarray}
 A_{C,vu} 
& = & \!\!\  \f{1}{2}\eth_C(A_{u,v})+\f{1}{4} \eth^B(\eth_B A_C-\eth_C A_B)\label{eq:ev_em}\;\;.
\end{eqnarray}
The set of equations can be solved provided the following initial data on $V_0$
\begin{subequations}
\begin{equation}
\label{ }
V_B(u,x^C) := A_{B,u}|_{v=0}\;\;,
\end{equation}
the data on $U_0$
\begin{equation}
\label{ }
U_B(v, x^C) := A_{B,v}|_{u=u_0}
\end{equation}
and the data on the common intersection $S_0$
\begin{eqnarray}
\label{ }
I(x^A) := A_{u,v}  |_{\stackrel{u=u_0}{v=v_0}}&,&
J(x^A):=A_{u}  |_{\stackrel{u=u_0}{v=v_0}}\nonumber \\&&\\
\tilde A_B(x^C)&:=& A_{B}|_{\stackrel{u=u_0}{v=v_0}}\;\;.\nonumber
\end{eqnarray}
\end{subequations}
With these initial data the Maxwell equations can be solved by a two-part algorithm:
\begin{enumerate}
  \item {\bf Part I } - {\it  solution on $V_0$ }
  
  \begin{enumerate}
  \item With the data $\tilde A_B$ on $S$ and the knowledge of $V_B$ on $V$, the value of $A_B$ is determined all along $V$ via the definition $\partial_u A_B |_{v=0}= V_B$.
  \item With the  supplementary equation \eqref{eq:suppl_em} the value of $\partial_v A_u$ is determined all along $V$ using the data $I$ and $V_B$.
\end{enumerate}
  \item {\bf Part II } - {\it  solution exterior to $V_0$  }
\begin{enumerate}
  \item   With the data $\tilde A_B$ on $S$ and the knowledge of $U_B$ on $U_0$, the value of $A_B$ is determined all along $U$ via the definition $\partial_v A_B = U_B$ for $u=u_0$.
  \item Given $I$ on $S$ and $U_A$ on V at the time $u=u_0$, $\partial_v A_u$ can be determined on $U$ for the time $u=u_0$ using the hypersurface equation \eqref{eq:hyp_em}.
  \item We use \eqref{eq:ev_em} to evolve $A_{C,u}$ from $u=u_0$ for $u=u_0+\Delta u$ where $0<\Delta u\ll 1$.
\end{enumerate}
\item{\bf repetition}
After completion of step three in part II,  all data are at hand to restart the (finite ) difference algorithm at step one in part I and calculate $A_B$ for $u=u_0+\Delta u$. 
\end{enumerate}
The above-described procedure to solve the Maxwell equations in null gauge, has  similarities to solving a nonlinear version of the non-linear wave  {equation} $\square \varphi = S$  \cite{CompleteNull}. 
In the latter case, first an advanced solution $\varphi_{+}$ of the homogeneous (source-free) wave equation is solved via some Greens function \cite{mathphys,jackson}. 
Then, the advanced solution is used  to find the retarded solution $\varphi_-$ by an integral over the source as a function of the advanced solution. 

For the nonlinear general relativistic equation of motion, a Greens function cannot be found for the retarded or advanced solutions. 
Instead, the `analogue' statement is to  {require  that} incoming or outgoing radiation vanishes (see. e.g. \cite{boostKS} where this is discussed in relation of boosted Kerr-Schild black holes in the Bondi-Sachs formalism). The electromagnetic counterpart of this would be to require that either $V_B$ vanishes on $v=0$ or $U_B$ vanishes on $u=0$.

The similarities of this electromagnetic example to the affine null metric formulation are the following:
\begin{itemize}
  \item [(i)] 
       The free initial data on the intersecting null hypersurfaces are the physically relevant propagating fields.  
      That is, if we label the with $x^\alpha = (x^0, x^1)$ the two coordinates varying along the  {rays} (corresponding to $x^\alpha = (u,v)$ for the metric \eqref{metric_dn_mink}), the physically propagating field are the angular components $A_B$ of the four potential and the data on the null hypersurfaces are $A_{B,\alpha}$. 
In the general relativistic case, the physical relevant degrees of freedom are the transverse-traceless degrees of freedom of the conformal 2-metric $h_{AB}$ 
(which encodes the gravitational waves). 
The free data on the null hypersurfaces are the shear $h_{AB, \alpha}$ of the respective initial data hypersurface. 
  \item [(ii)] The inital values of the physical propagating fields are only prescribed on the common intersection of the pair of null hypersurfaces and depend only on angular coordinates $x^A$, i.e. $A_B$ for the electromagnetic case and $h_{AB}$ for the general relativistic case. 
  \item 
  [(iii)] Additional variables having lower tensor rank than the propagating fields are prescribed on the common intersection. 
These fields include derivatives wrt $x^\alpha$ and depend only on angular coordinates $x^A$. 
For the electromagnetic case, where the propagating field has tensor rank 1, these additional fields are of rank 0, namely the scalars $I$ and $J$. 
In the general relativistic case, these fields are scalar fields (rank 0) and a 2-vector field (rank 1) on the  common intersection, which are in general derivatives with respect to $x^\alpha$ and also depend only on $x^A$. 

\end{itemize}

\section{Affine null formulation}\label{sec:dn_system}
In a four dimensional (at least three times) differentiable Lorentz manifold $(\mc{M}^4,\, g_{ab})$ consider two distinct null hypersurfaces $\mc{B}$ and $\mc{N}_0$ whose common intersection is $\Sigma = \mc{B}\bigcap \mc{N}_0$. 
The intersection $\Sigma$ is a two dimensional submanifold of $\mc{M}^4$. In $\Sigma$, we can always choose two coordinates $x^A$ such that the intrinsic metric $g_{AB}$ of $\Sigma$ obeys the conformal decomposition (see \cite{StewardBondiMass} for a related discussion) $$g_{AB} = r^2h_{AB}.$$
On any point in $\Sigma$, we choose two null vectors $k^a_\Sigma=k^a_\Sigma(x^A)$ and $n^a_\Sigma = n^a_\Sigma(x^A)$ with the properties 
i) $k^a_\Sigma$ is tangent to $\mc{N}_0$, 
ii) $n^a_\Sigma$ is tangent to $\mc{B}$, 
iii) $g_{ab}k^a_\Sigma n^b_{\Sigma} = -1$ on $\Sigma$ and that 
iv) both $k^a_\Sigma$ and $n^a_\Sigma$ are orthogonal to coordinate directions $\partial_A$ in $\Sigma$. 
Condition (iv) also means  the two coordinates $x^A$ obey $k^a_\Sigma x^A_{,a}  = n^a_\Sigma x^A_{,a} = 0$.
Since $k^a_\Sigma$ and $n^a_\Sigma$ are null vectors they are also orthogonal to their respective null hypersurface. 
A further consequence of the null property of $k^a_\Sigma$ and $n^a_\Sigma$ is that they are tangent vectors of null geodesics $k^b_\Sigma\nabla_b k^a_\Sigma = \Phi_{1}k^a_\Sigma $ and $n^b_\Sigma\nabla_b n^a_\Sigma = \Phi_{2}n^a_\Sigma$   emanating from $\Sigma$, where $\Phi_1$ and $\Phi_2$ are arbitrary scalars depending on the coordinates. 
These null geodesics are called rays.  

Let $\lambda$ be an affine parameter for the rays with the tangent vector $k^a_\Sigma$ with $\lambda=0$ on $\Sigma$. 
As the null geodesics are affinely parameterised, $\Phi_1=0$. 
The solution of the geodesic equation   $k^b_\Sigma\nabla_b k^a_\Sigma=0 $ with  $k^a_\Sigma=dx^a/d\lambda|_\Sigma$ and $\lambda=0$ on $\Sigma$ generates a spray of null rays 
emanating from $\Sigma$.
These null rays form the hypersurface $\mc{N}_0$ and we carry over the definition $k^a_\Sigma=dx^a/d\lambda|_\Sigma $ at $\Sigma$ to every point in $\mc{N}_0$  so that $k^a(\lambda, x^A) = dx^a/d\lambda|_{\mc{N}_0} $.  
We also require that the conditions $k^a_\Sigma x^A_{,a}  = 0$ carry  over  to every point in  $\mc{N}_0$, so that $k^ax^A_{,a} = 0$  on   $\mc{N}_0$.
These two conditions also imply $g_{\lambda\lambda}=g_{\lambda A} = 0$ on  {$\mc{N}_0$.}
Since $\Sigma$ is the intersection of $\mc{N}_0$ and $\mc{B}$, the null hypersurface $\mc{B}$ is represented by $\lambda=0$ on $\mc{N}_0$.

Let the scalar function $w$ be the collection of points in $\mc{N}_0$ ran through by the ray congruence $k^a$ with start vectors $k^a_\Sigma$ on $\Sigma$, we than have  $k^aw_{,a} = 0$, i.e. $w=const$ as $k^a$ varies. 
On $\Sigma$, we  choose $w$ as  affine parameter for the null rays with initial start vector $n^a_\Sigma$ , then $\Phi_2=0$.
With $n^a_\Sigma = dx^a/dw|_{\Sigma}$ and $w=0$ on $\Sigma$, the solutions of the geodesic equation $n^b_\Sigma\nabla_b n^a_\Sigma=0 $ are a spray of null geodesics generating $\mc{B}$. 
Proceeding similarly as for $\mc{N}_0$, the conditions $n^a_{\Sigma}x^a_{,a}=$ and $n^a_\Sigma = dx^a/dw|_{\Sigma}$ can be carried for every point in $\mc{B}$ so that $n^a(w,x^A)= dx^a/dw|_{\mc{B}}$ and $n^ax^A_{,a}=0$ on $\mc{B}$ giving $g_{ww}=g_{wA} = 0$ on $\mc{B}$. 

Up until now the three dimensional intrinsic coordinates $(\lambda, x^A)$ and $(w, x^A)$ to both null hypersurfaces have been set up only. 
Let $\mc{P}$ be an arbitrary point in the neighbourhood of $\Sigma$ being neither on $\mc{N}$ nor on $\mc{B}$. 
To reach $\mc{P}$ from $\mc{N}$ (or $\mc{B}$), we make first the observation that at any point on any given cross section $\Sigma^\prime$ of $\mc{N}_0$ at $\lambda=\lambda^\prime>0$ ( or of $\mc{B}$ at $w=w^\prime>0$) there is a  point $\mc{P}^\prime$  with a unique null vector $\ell^a$ orthogonal to $k^a$ (or $n^a$)  obeying the orthogonality condition $\ell_a k^a=-1$ (or $\ell_a n^a=-1$) at $\mc{P}^\prime$.
The vector $\ell^a$ is the tangent vector of a null geodesic $\overrightarrow{\mc{P}^\prime\mc{P}}$ starting at $\mc{P}^\prime$ and connecting any point $\mc{P}$ in the neighbourhood of the null hypersurface on which is $\mc{P}^\prime$.
Since $\mc{P}^\prime$ is on either $\mc{N}_0$ or $\mc{B}$, it can be connected with a unique geodesic emanating from a point $\mc{P}_0\in \Sigma$ with either a given tangent vector $k^a_{\mc{P}_0}$ at $\mc{P}_0$ if $\mc{P}^\prime\in \mc{N}$ or with tangent vector $n^a_{\mc{P}_0}$ at $\mc{P}_0$ if $\mc{P}^\prime\in \mc{B}$. 
If $\mc{P}^\prime \in \mc{N}_0$ we parallel transport  $n^a_{\mc{P}_0}\rightarrow n^a_{\mc{P}^\prime}$ to $\mc{P}^\prime$
while  if $\mc{P}^\prime \in \mc{B}$ the vector $k^a_{\mc{P}_0}\rightarrow k^a_{\mc{P}^\prime}$ is parallel transported to $\mc{P}^\prime$.
Depending if $\mc{P}^\prime\in \mc{N}_0$ or $\mc{P}^\prime\in \mc{B}$, the  vector $\ell^a$ will have the respective relation $\ell^a\propto n^a_{\mc{P}^\prime}$ on $\mc{N}$ or $\ell^a\propto k^a_{\mc{P}^\prime}$  on $\mc{B}$, because both vectors obey the same normalisation condition  with respect to the tangent vector field of the geodesic along which the parallel transport had been performed. 
It is seen that any point not on the null hypersurfaces can be reached in a similar way from the common intersection. 

The null vectors $k^a$ and $n^a$  define their respective one-forms by 
\begin{equation}
\label{ }
k_a|_\mc{N} = \epsilon w_{,a}|_\mc{N} \;\;,\;\; n_a|_\mc{B} = - \lambda_{,a}|_\mc{B} \;\;,
\end{equation}
on $\mc{N}$and  $\mc{B}$, respectively.  Where, in particular, $w\ge 0$ and the parameter $\epsilon$ indicates whether $w=const$ is an ingoing null hypersurface  ($\epsilon = 1$) or outgoing null hypersurface ($\epsilon = -1$) as seen from a timelike observer on $\Sigma$.

We are now in the stage to complete the four dimensional  coordinate chart  $x^a$ for the affine, null-metric initial value formulation consisting two intersecting null hypersurfaces and their common intersection.
We refer to one of these null hypersurfaces as null boundary $\mc{B}$ and the other one as initial data surfaces $\mc{N}_0$.
The null boundary may be for example a stationary or dynamical horizon, and the initial data surface is the null hypersurface constant to ingoing or outgoing radiation from such a horizon.
The two coordinates $x^A$  parametrise the intersection  $\Sigma$ according to their previous definition. 
The additional  two coordinates, $x^0=w$ and $x^1=\lambda$,   are  the parameters along the null rays generating $\mc{N}_0$  and $\mc{B}$, respectively. 
Note, given the previous description, either of the two null hypersurfaces $\mc{N}_0$ or $\mc{B}$ may be the boundary surface or initial data surface because any point not being on these two hypersurfaces can be reached in a similar way.
We choose $\mc{N}_0$ corresponding to the three dimensional hypersurface $w=0$ as the initial dats surface and $\mc{B}$  as the boundary represented by the null hypersurface intersecting $\mc{N}_w$ at $\lambda=0$. 
By this construction, $w$ labels distinct null hypersurfaces $\mc{N}_w$ emanating from $\mc{B}$.
For each value of $w$, there is a common distinct intersection $\Sigma_w$ between $\mc{B}$ and $\mc{N}_w$.
As a consequence we require $k^a w_{,a} = k^a x^A_{,a}=0$ to hold everywhere since the null rays with the tangent  {vector field} $k^a$ emanate from $\mc{B}$ so that 
\begin{subequations}\label{coordinate_cond}
\begin{equation}
\label{eq:fam}
g_{\lambda\lambda}=g_{\lambda A}=0\;\mbox{everywhere.}
\end{equation}
Moreover, from choice of $w, x^A$  as intrinsic  {coordinates} on $\mc{B}$ there is 
\begin{equation}
\label{eq:bound}
g_{ww}=g_{wA}=0\mbox{ on }\mc{B}.
\end{equation}
Equations \eqref{coordinate_cond} are the coordinate  representation of the choice $\mc{B}$  as boundary and $\mc{N}$ as family of null hypersurfaces emanating from $\mc{B}$.
Following  \cite{CompleteNull,Win_affine}, we choose $w$ as an affine parameter along the null rays generating $\mc{B}$, then evaluation of the  geodesic equation $n^a\nabla_a n^b=0$ on $\mc{B}$ implies 
$$g_{ww,\lambda}=0\;\; \mbox{on}\;\; \mc{B}.$$
We require  the normalisation $k^an_a = -1$ everywhere so that 
\begin{equation}
\label{ }
 {g_{w\lambda}=\epsilon}\;\mbox{everywhere.}
\end{equation}
\end{subequations}
We further require $\lambda\ge 0$  and   $|\epsilon| =1$ via the normalisation $k^a\partial_a \lambda = \epsilon$ on $\mc{B}$ meaning the geodesic rays with tangent vectors $k^a$ are outgoing ($\epsilon=-1$)  from $\mc{B}$  or ingoing $\epsilon=1$ towards  $\mc{B}$.
The line element resulting from the conditions \eqref{coordinate_cond} is of the form \cite{CompleteNull,Win_affine}
\begin{eqnarray}
\label{eq:dn_metric}
ds^2 &=& -W dw^2 +2\epsilon dwd\lambda\nonumber\\
&&+ r^2h_{AB}(dx^A-W^Adw)(dx^B-W^Bdw)\;\;,\;\\\
0&=&W|_{\lambda=0}=W_{,\lambda}|_{\lambda=0}=W^A|_{\lambda=0} = \epsilon^2-1
\end{eqnarray}
with 
$W, r, W^A, h_{AB}$ being functions of $x^a=(w,\lambda, x^A)$ and $$h^{AC}h_{CB} = \kron{A}{B}\;, \;\det(h_{AB})=h(x^A)$$
as well as indices $x^A$ are raised (lowered) with $h^{AB}$ ($h_{AB}$), e.g. $W_A = h_{AB}W^A$. 
The covariant derivative with respect to $h_{AB}$ is denoted with $\mc{D}_A$. 
Note that $h_{AB}$ has only 2 degrees of freedom, because of the determinant condition. 
A suitable parameterisation for $h_{AB}$ in terms of standard spherical coordinates $x^A = (\theta, \phi)$ is \cite{vdB,BSscolar}
\begin{eqnarray}
h_{AB}dx^Adx^B & = & \cosh(2\delta)\Big[e^{2\gamma}d\theta^2 + e^{-2\gamma}\sin^2\theta d\phi^2\Big]\nonumber \\
 &  & +2\sin\theta \sinh(2\delta)d\theta d\phi\;\;.
\end{eqnarray}
The inverse metric has the nonzero components
\begin{equation}
\label{ }
g^{\lambda w }=\epsilon\;,\;
g^{\lambda \lambda}=W\;,\;
g^{\lambda A}= \epsilon W^A\;,\;
g^{AB} = \f{1}{r^2}h^{AB}\;\;.
\end{equation}
The null vectors $k^a$ and $n^a$ have the coordinate expressions
\begin{eqnarray}
k^a\partial_a& = & \partial_\lambda \;\;,\\
n^a\partial_a & = & -\epsilon \partial_w -\f{1}{2}W\partial_\lambda -\epsilon W^A\partial_A\;\;.
\end{eqnarray}
where in particular $n^a\partial_a|_{\mc{B}} =-\epsilon \partial_w$.
The expansion rates, $\theta_{(k)}:= \nabla_ak^a$ and $\theta_{(n)} := \nabla_an^a$, for both null vectors are
\begin{eqnarray}
\label{ }
\theta_{(k)} &=&\partial_\lambda \ln r^2 \\
\theta_{(n)} &=& -\epsilon \partial_w \ln r^2  {- \f{(r^2 W)_{,\lambda}}{2r^2}}-\f{\epsilon \mc{D}_A (r^2 W^A)}{r^2}\;\;.
\end{eqnarray}
Observe, that $\theta_{(k)}$ and $\theta_{(n)}$ depend on $\mc{B}$ only on the conformal factor $r$.
We also introduce a complex dyad $m^A$ to represent the conformal 2-metric $h_{AB} = m_{(A}\bar m_{B)}$, where $m^Am_A = m_A\bar m^A-1 = 0$. 
\subsection{Infinitesimal transformations on the boundary $\mc{B}$ }\label{sec:trafos}
On a given initial data surface  {$\mc{N}_w$}, $w=const$,  consider the infinitesimal coordinate transformations $x^a\rightarrow x^a+\xi^a$ which should leave the structure of the geometry on the initial null hypersurface at $\lambda=0$ invariant. Consequently they must comply with the  conditions
\begin{equation}
\label{eq:cond_Lie_g0a}
\mathcal{L}_\xi g^{wa} =0\;\;\;
\end{equation}
where $\mc{L}_\xi$ is the Lie derivative
$$
\mc{L}_\xi g^{ab} = \xi^c g^{ab}_{\p{ab},c} - g^{ac}\xi^b_{\p{c},c}-g^{cb}\xi^a_{\p{c},c}\;\;.
$$
Furthermore,  we wish to preserve the behaviour of the  metric functions $W, W^A, g^{AB}$ on the boundary so that 
\begin{equation}
\label{eq:Lie_cond_other}
\mathcal{L}_\xi g^{\lambda\lambda}  = O(\lambda^2) \;\;,\;\;
\mathcal{L}_\xi g^{\lambda A} = O(\lambda)\;\;,\;\;
\mathcal{L}_\xi g^{A B}  = O(\lambda)\;\;,
\end{equation}
We assume hereafter that the non-zero components of  metric have the following regular expansion in term of the affine parameter 
\begin{eqnarray}
r&=&r_{(0)}(w, x^A)+ r_{(1)}(w,x^A)\lambda+ O(\lambda^2)\\
g^{w\lambda}&=&\epsilon\\
g^{\lambda\lambda} & = & W_{(2)}(w,x^A)\lambda^2 +... \\
g^{\lambda A} & = &\epsilon W^A_{(1)}(w,x^B)\lambda +... \\
h^{A B} & = & h^{AB}_{(0)}(w,x^C)+h^{AB}_{(1)}(w,x^C)\lambda +O(\lambda^2)
\end{eqnarray}
and where the orthogonality condition $h_{AB}h^{BC}=\kron{C}{A}$  and determinant condition $\det(h_{AB}) = h(x^A)$ imply
\begin{eqnarray}
\kron{A}{C}&=&h^{AB}_{(0)}h_{(0)BC}\;, \\
h_{(1)AB}&:=&h_{(0)AE}h_{(0)AF}h^{EF}_{(1)}\;,\\
h_{A B} & = & h_{(0)AB}-h_{(1)AB}\lambda + O(\lambda^2)\;, \\
0&=& h_{(0)EF}h^{EF}_{(1)}\;.
\end{eqnarray}
Conditions \eqref{eq:cond_Lie_g0a} give the solution
 {\begin{eqnarray}
\label{ }
\xi^a\partial_a &=&\alpha(w, x^A)\partial_w
+ \Big[f^A(w, x^C)+ O^A(\lambda)\Big]\partial_A  \nonumber\\
&&
\!\!\!+\Big[\beta_{(0)}(w, x^A) + \beta_{(1)}(w, x^A)\lambda+O(\lambda^2)\Big]\partial_\lambda\;\;.
\end{eqnarray}
}
Calculation of  \eqref{eq:Lie_cond_other} and $g_{AB}\mc{L}_\xi  g^{AB} = 0$ gives to leading order in the $\lambda$ expansion
\begin{eqnarray}
0 & = &  {\beta_{(0),w}} \label{eq:beta_eqn}\\
0 & = & f^A_{,w}- {\beta_{(0)}} W^A_{(1)}+\epsilon \f{h^{AB}_{(1)}}{r^2_{(0)}} {\beta_{(0), B}}\label{eq:fA_eqn}\\
0 &=&  \Big[\alpha ( \ln r_{(0)}^2)_{,w} + {\beta_{(0)}}\Big(\f{r_{(1)}}{r_{(0)}}\Big)+ ( \ln r_{(0)}^2)_{,C}f^C\Big]h^{AB}_{(0)} 
\nonumber\\&& +2\;_0\mc{D}^{(A}f^{B)}-\alpha [h^{AB}_{(0)}]_{,w} -  {\beta_{(0)}} h^{AB}_{(1)} \label{eq:DAfB_eqn}\\
0 &=&  \Big[\alpha ( \ln r_{(0)}^2)_{,w} + {\beta_{(0)}}\Big(\f{r_{(1)}}{r_{(0)}}\Big)+ ( \ln r_{(0)}^2)_{,C}f^C\Big]
\nonumber\\&& +\;_0\mc{D}_{B}f^{B}  \label{eq:DAfA_eqn}
\end{eqnarray}
where $\;_0\mc{D}_{A}$ is the covariant derivative with respect to $h_{(0)AB}$.
Eq.~\eqref{eq:beta_eqn}   implies  { $\beta_{(0)} = \beta_{(0)}(x^A)$. }
Then, given an expression for  {$\beta_{(0)}$,}  we can solve \eqref{eq:fA_eqn} for $f^A(w,x^C)$. 
The resulting solution will depend on a free function $f^A_0(x^B)$  and the (in general non zero) values of $r_{(0)}$, $W^A_{(1)}$, $h^{AB}_{(1)}$. 
We make the requirement that infinitesimal coordinate transformations should not depend on these dynamical fields  $r_{(0)}$, $W^A_{(1)}$, $h^{AB}_{(1)}$ \cite{Donnay2}. 
This implies $ {\beta_{(0)}}=0$ and thus $f^A_{,w}=0$ so that $f^A = f^A(x^B)$. 
 Moreover, considering the $O(\lambda)$ term of $\mc{L}_\xi g^{w\lambda}$ gives us {$\beta_{(1)} = -\alpha_{,w}$}. Next, calculation of the  $\mc{L}_\xi g^{\lambda\lambda}=O(\lambda^2)$ provides  {$\beta_{(1),w}=0$}, therefore we have $\alpha_{,ww}=0$. The latter condition was also found in a coordinate independent calculation by \cite{curvature_alpha}. 
Manipulation of  \eqref{eq:DAfB_eqn} and \eqref{eq:DAfA_eqn} gives us
\begin{eqnarray}
\!_0\mc{D}_{(A}f_{B)} - \f{1}{2}h_{(0)AB}\Big[ \!_0\mc{D}_{E}f^{E}\Big]&=&\f{\alpha}{2}  h_{(0)AB,w}    \;\;.
\end{eqnarray}
This relation shows that if the conformal metric $h_{AB}$ on $\mc{B}$ does not vary along the generators $\partial_w$ (i.e.  $\mc{B}$ is shear free null hypersurface) then $f^A$ is a conformal Killing vector.

The most general allowed coordinate transformations $x^a\rightarrow x^a+\xi^a$ on the boundary $\mc{B}$ given our assumptions are
\begin{eqnarray}
\label{ }
\xi^a\partial_a|_{\mc{B}} &=&\alpha(w, x^A)\partial_w +f^A(x^C)\partial_A\;,\;
\alpha_{,ww}=0.\qquad
\end{eqnarray}
Here $\alpha$ is referred to as the BMS-type supertranslation in the context of  $\mc{B}$ being a horizon \cite{Donnay1, Donnay2}. 
We prefer to the notation BMS-type supertranslation for $\alpha(w,x^A)$, because a BMS supertranslation (as it is found at null infinity \cite{Bondi,Sachs}) is function depending {\it only} on $x^A$ and {\it not} on the three parameters $(w, x^A)$. 
The additional dependence in $w$ of $\xi^w$, arises because $\mc{B}$ is a general null hypersurface that does not have any restrictions, as e.g. given by an asymptotical fall off.  
However, if we  require that the null normal $n^a$ is preserved along $\mc{B}$, $\mc{L}_\xi n^a = 0$, we have $\xi^w_{,w} = 0$ \cite{Blau}  implying
\begin{equation}
\label{eq:BMSalpha}
\alpha(w, x^A)\rightarrow \alpha_{BMS}(x^A)
\end{equation}
which is a proper supertranslation as known from the Bondi-Sachs work \cite{BSscolar}. 

\subsection{Field equations}\label{sec:EE}
We  {consider} the vacuum Einstein equations $R_{ab}=0$, where  $R_{ab}$ is the Ricci tensor. 
As shown by Sachs \cite{Sachs_civp} and also in App.~\ref{app:gen_null}, the  Einstein equations for \eqref{eq:dn_metric} can be grouped into three supplementary equations $R_{ww}=R_{wA}=0$ on $\mc{B}$, six main equations $R_{\lambda\lambda}=R_{\lambda A}=g^{AB}R_{AB}=m^Am^BR_{AB}=0$ and one trivial equation $R_{w\lambda}=0$. 
The twice contracted Bianchi identities imply that if the main equations hold on one null hypersurface $\mc{N}_w$,  the supplementary equations hold everywhere on that surface  if they hold at on $\mc{B}$. 
In addition to that, the trivial equation is an algebraic consequence of the main equations. 
The calculation of the relevant Ricci tensor components is rather tedious and details of this calculation starting out from a  most general  metric \eqref{eq:gen_null} at a null hypersurface are displayed in App.~\ref{sec_genRic} and App.~\ref{sec_confRic}.
The main equations consist of  three  hypersurface equations:  one  equation  for $r$ along the rays generating $\mc{N}$ \footnote{This equation also corrects a typo in the respective equation in \cite{Win_affine,CompleteNull}}
\begin{subequations}\label{eq:hyp_EE}
\begin{eqnarray}
0 & = &   r _{,\lambda\lambda}
	+ \f{r}{8} h_{CB,\lambda}h_{DA,\lambda}h^{AC}h^{BD}
\label{eq:hyp_r}
\end{eqnarray}
and two equations allowing to determine the shifts $W^A$
\begin{eqnarray}
0&=&
	-\f{\epsilon}{2r^2}\Big(r^4  h_{AB}W^B_{,\lambda}\Big)_{,\lambda}
	 {-\mc{D}_A	(\ln r)_{,\lambda}	}
	+  {\f{1}{2r^2}\mathcal{D}^B (r^2h_{AB,\lambda}) \;\;.}
\nonumber\\&&
	\label{eq:hyp_WA}
\end{eqnarray}
\end{subequations}
Note that these equations do not contain derivatives with respect to $w$.

The three remaining main equations determine  the mixed $\partial_w\partial_\lambda$ derivatives of the metric $g_{AB}$.
There is  one for equation for $r$ and two for  the conformal 2-metric  $h_{AB}$ 
\begin{subequations}\label{eq:ev_EE}
\begin{eqnarray}
0
 &=&
 \epsilon\mathcal{R} 
- \epsilon \mc{D}^F\mc{D}_F \ln r^2 
	-\Big[2  (r^2 )_{,w} 	+\epsilon W(r^2 )_{,\lambda}\Big]_{,\lambda}\nonumber\\
&&	
	-\mc{D}_C\Big[\f{1}{r^{2}} ( r^4 W^{C})_{,\lambda}\Big]
	-\f{\epsilon}{2}r^4 h_{EF}	W^E_{,\lambda}W^F_{,\lambda}	
\label{eq:ev_r}
\end{eqnarray}
\begin{widetext}
\begin{eqnarray}
0
 &=& m^Am^B\Bigg\{
	 r(rh_{AB})_{,w\lambda}
 	 +\f{\epsilon}{2}\Big[r^2W h_{AB,\lambda}\Big]_{,\lambda} 
	  +r^2W^{C}\mc{D}_Ch_{AB,\lambda}
	  +\f{1}{2}\Big[\mc{D}_C(r^2 W^{C})\Big] h_{AB,\lambda}
	+( r^2)_{,\lambda}  h_{BC} \mc{D}_A W^C  
\nonumber\\&&
\qquad	\qquad\;\;	
	+r^2h_{EA} \mc{D}_{B}  W^E_{,\lambda}
	-r^2h_{FA}h_{BC,\lambda} \Big(\mc{D}^C W^F -  \mc{D}^F W^C   \Big)   
	+\f{\epsilon r^4}{2} h_{EA}h_{BF}  W^E_{,\lambda}W^F_{,\lambda}
	\Bigg\}\label{eq:ev_hAB}
\end{eqnarray}
\end{widetext}
\end{subequations}
where $ \mathcal{R}$ is the Ricciscalar of $h_{AB}$. 
The supplementary  equations   on $\mc{B}$ are 
\begin{subequations}
\begin{equation}
\label{eq:supp_r}
   \f{r_{,ww}}{r}  
	 =
	  -\f{1}{8} h_{CB,w}h_{DA,w}h^{AC}h^{BD}
\end{equation}
 and two equations for $W_{A,\lambda}$ on $\mc{B}$
\begin{eqnarray}
\epsilon [  r^4h_{AB}W^B_{,\lambda}]_{,w}
 &=&
 { r^2\mc{D}_A (\ln r^2)_{,w} }
-\mc{D}^B(r^2 h_{AB,w})
	\nonumber\\
\end{eqnarray}
\end{subequations}
Evaluation of the evolution equations on $\mc{B}$ gives
\begin{subequations}
 \begin{eqnarray}
0
 &=&
\epsilon\mathcal{R} 
- \epsilon \mc{D}^F\mc{D}_F \ln r^2  
	-2  (r^2 )_{,w\lambda}
	-\mc{D}_C( r^2 W^{C}_{,\lambda})
\nonumber\\&&
	-\f{\epsilon}{2}r^4 h_{EF}	W^E_{,\lambda}W^F_{,\lambda}	\label{eq:ev_r_B}
\end{eqnarray}
and
\begin{eqnarray}
0
 &=& 
	m^Am^B \Big[r(rh_{AB})_{,w\lambda} \Big]
	 {+\f{r^2}{2}m_E m^B D_{B}  W^E_{,\lambda}}
\nonumber\\&&
	 - \f{\epsilon r^4}{2}  \Big(m_ E W^E_{,\lambda}\Big)^2
	 \label{eq:bound_ev_hAB_ulambda_intro}
\end{eqnarray}
\end{subequations}
We also will make use of the  complex Weyl scalar
\begin{eqnarray}\label{psi2}
\Psi_2&=&-C_{abcd}k^aM^b\bar M^cn^d \;\;.
\end{eqnarray} 
where $M^a\partial_a =  r^{-1}m^A\partial_A $ and $C_{abcd}$ is the Weyl scalar. As we consider vacuum spacetimes with $R_{ab}=0$, the Weyl tensor is equal to the Riemann tensor $R_{abcd}$ since
$$
C_{abcd} = R_{abcd} - R_{a[c}g_{d]b} + R_{b[c}g_{d]a} +\f{R}{3}g_{a[c}g_{d]b} \;\;,
$$
where $R = R^a_{\p{a}a}$ is the Ricci scalar with respect to $g_{ab}$.

\section{A hierarchical set of equations on $\mc{B}$ and $\mc{N}$}\label{sec:hierarchy}
\subsection{Definition of new variables}
The original equations of the affine null formulation \eqref{eq:hyp_r}-\eqref{eq:bound_ev_hAB_ulambda_intro} are not hierarchical, because
the evolution system \eqref{eq:ev_EE}  is coupled with the hypersurface equations \eqref{eq:hyp_EE}.
Winicour  \cite{Win_affine} decoupled these equations for the timelike-null  initial value formulation. 
Here we  follow closely his approach with some generalisations and the necessary adaptation regarding the boundary $\mc{B}$. 
Like in \cite{Win_affine} , we introduce the variables
\begin{eqnarray}
\label{eq:def_YJ}
Y &=& W +\epsilon \f{2 r_{,w}}{r_{,\lambda}}
\\
J &=&  r_{,\lambda}( h_{AB,w} m^Am^B)
	 -r_{,w}  (h_{AB,\lambda}m^Am^B)\label{eq:def_J}
\end{eqnarray}
as well as  first derivatives of conformal factor $r$ and the conformal 2-metric  $h_{AB}$
\begin{equation}
\label{eq:def_rho_theta_n_sigma}
 \rho=r_{,w}\;\;,\;\;
  {\Theta = r_{,\lambda}\;\; ,\;\;n_{AB} = h_{AB,w}}
\;\;,\;\;\sigma_{AB} = h_{AB,\lambda}
\end{equation}together with the mixed derivative 
\begin{equation}
\label{ }
\mu:= r_{,w\lambda} = \rho_{,\lambda}=\Theta_{,w}\;\;.
\end{equation}
The fields $n_{AB}$ and $\sigma_{AB}$ are most conveniently  expressed as 
\begin{eqnarray}
n_{AB} &=&\nu \bar m_A  \bar m_B + \bar \nu m_A m_B \;\;,\;\;\label{eq:decomp_n}\\
\sigma_{AB} &=&  \sigma \bar m_A  \bar m_B + \bar \sigma m_A m_B \label{eq:decomp_sigma}
\end{eqnarray}where $\nu$ and $\sigma$ are the spin weight 2 scalars defined as 
$\sigma =  m^Am^B\sigma_{AB}$ and $\nu =  m^Am^B n_{AB}.$
The new variable $J$ is   { a spin} weight 2 field,
$J =  \Theta\nu 
	 -\rho  \sigma.$
The variable $\Theta$  relates to the expansion of the null rays $\nabla_a k^a$ on the families of null hypersurfaces $\mc{N}$ while $\rho$ relates to the expansion $\nabla_a n^a$ of the null rays on $\mc{B}$.

For indices $i,j\in\{w, \lambda\}$ and using common relation of a derivative of a determinant
$$h h^{CD}h_{CD,i} = h_{,i}$$ 
there is the useful relation 
\begin{widetext}
\begin{eqnarray}
h_{AB,i}h_{CD,j}h^{AC}h^{BD} 
& = &  (\bar m^A \bar m^B h_{AB,i})(m^Cm^Dh_{CD,j})+(m^A m^Bh_{AB,i})(\bar m^C\bar m^D h_{CD,j})\;\;.
 \end{eqnarray}
 \end{widetext}
 {This gives us}
\begin{eqnarray}
h_{AB,w}h_{CD,w}h^{AC}h^{BD} & = & 2\nu \bar \nu  \label{eq:reformhAB,0hAB,0}\\
h_{AB,\lambda}h_{CD,\lambda}h^{AC}h^{BD} & = & 2\sigma \bar \sigma \label{eq:reformhAB,1hAB,1}
\end{eqnarray}
also note \footnote{Since $h_{AB} =2 m_{(A}\bar m_{B)}$ with $m^Am_A = 0$ implies $m^Am^Bh_{AB} = 0$ and $m^A\bar  m_{A,\lambda}=0$ so that ${m^Am^B_{,\lambda}h_{AB,c} = [(m^Am_{A,c})(m^B_{,\lambda}\bar m_B)+(m^A\bar m_{A,\lambda})( m^B_{,c} m_B)]=0}$.\label{fn:DmAmADhAB}}
\begin{equation}
\label{eq:nu,1 = sigma,0}
m^A m^B h_{AB,w\lambda} = \nu_{,\lambda} = \sigma_{,w}
\end{equation}

\subsection{Reformulation of the hypersurface equations}
This section is a review and  independent reproduction of the principal results of  \cite{Win_affine}. 
With \eqref{eq:reformhAB,1hAB,1}, the hypersurface equation \eqref{eq:hyp_r} becomes
\begin{equation}
\label{eq:hypersurf_r_sig}
\f{ r _{,\lambda\lambda}}{r} = 
	- \f{1}{4}\sigma\bar \sigma
\end{equation}
The new variable $Y$  allows us to write the evolution  equation \eqref{eq:ev_r} as an hypersurface equation for $Y$
\begin{equation}
\label{eq:Y}
 	\Big[Y(r^2)_{,\lambda} \Big]_{,\lambda}
	=
	\mathcal{R}
	  + F_{[Y_{,\lambda}]}(r,h_{AB}, W^A)
\end{equation}
with 
\begin{eqnarray}
\label{ }
\lefteqn{F_{[Y_{,\lambda}]}(r,h_{AB}, W^A )
=	 
- \mc{D}^F\mc{D}_F \ln r^2 }
\nonumber\\&&
	-\epsilon\mc{D}_C\Big[\f{1}{r^{2}} ( r^4 W^{C})_{,\lambda}\Big]
	-\f{1}{2}r^4 h_{EF}	W^E_{,\lambda}W^F_{,\lambda}		
\end{eqnarray}
which vanishes in spherical symmetry. 
As outlined in \cite{Win_affine},	
the principal part of the evolution equation \eqref{eq:ev_hAB} for $h_{AB}$ can be written as  {(also see \eqref{eq:nu,1 = sigma,0} and the related footnote)}
\begin{eqnarray}
\lefteqn{m^Am^B\Big[
	 r(rh_{AB})_{,w\lambda}
 	 +\f{\epsilon}{2}(r^2Wh_{AB,\lambda})_{,\lambda} \Big] }\nonumber\\
&&	 = r \Big(\f{r J}{r_{,\lambda}}\Big)_{,\lambda}
 	 +\f{\epsilon}{2} \Big[r^2Y\sigma\Big]_{,\lambda}  
\label{eq:prinpart_ev_hAB}
\end{eqnarray}
Definition of 
\begin{eqnarray}
\lefteqn{F_{[\chi_{,\lambda}]}(h_{AB}, W^A, W)
 = m^Am^B \Bigg\{
 	 \f{\epsilon}{2}\Big[r^2Yh_{AB,\lambda}\Big]_{,\lambda} }
 \nonumber\\
&	  +r^2W^{C}\mc{D}_Ch_{AB,\lambda}
	  +\f{1}{2}\Big[\mc{D}_C(r^2 W^{C})\Big] h_{AB,\lambda}
\nonumber\\
&	+( r^2)_{,\lambda}  h_{BC} \mc{D}_A W^C  
	+r^2h_{EA} \mc{D}_{B}  W^E_{,\lambda}
\nonumber\\
&	
	-r^2h_{FA}h_{BC,\lambda} \Big(\mc{D}^C W^F -  \mc{D}^F W^C   \Big)   
\nonumber\\
&	+\f{ \epsilon r^4}{2} h_{EA}h_{BF}  W^E_{,\lambda}W^F_{,\lambda}
	\Bigg\}
	\label{eq:F_J}
\end{eqnarray}
and substitution of \eqref{eq:prinpart_ev_hAB} and \eqref{eq:F_J} into  \eqref{eq:ev_hAB} gives an hypersurface equation 
\begin{equation}
- r \chi_{,\lambda}
 =F_{[\chi_{,\lambda}]}(h_{AB}, W^A,  W^A_{,\lambda}, Y)\;\;.
\end{equation}
for the auxiliary variable 
\begin{equation}
\label{eq:def_chi}
\chi:=\f{r J}{\Theta}\;\;.
\end{equation}
The new variable $\chi$ determines the field $J$ from known fields $r$ and $\Theta$.
The variable $\rho$ needs to be determined by a hypersurface equation and this equation is obtained by taking the time derivative of \eqref{eq:hypersurf_r_sig}
 \begin{eqnarray}
0&=&
  \rho _{,\lambda\lambda }
	+\f{1}{4}  \rho \sigma\bar \sigma
	+\f{1}{4} r( \sigma\bar \sigma_{,w}
	+\sigma_{,w}\bar \sigma)
\end{eqnarray}
 {As of \eqref{eq:nu,1 = sigma,0},} we have
$$(m^Am^Bh_{AB,\lambda})_{,w} =(m^Am^Bh_{AB,w})_{,\lambda}\;\;, $$ so that
\begin{equation}
\label{eq:hyp_rho_0}
0=	  \rho _{,\lambda\lambda }
	+\f{1}{4}  \rho \sigma\bar \sigma
	+\f{1}{4} r( \sigma\bar \nu_{,\lambda}
	+\nu_{,\lambda}\bar \sigma)\;\;.
\end{equation}
The derivative $r\nu_{,\lambda}$ can be eliminated in \eqref{eq:hyp_rho_0} using the definition  \eqref{eq:def_J} of $J$  in \eqref{eq:hyp_rho_0}
so that
\begin{eqnarray}
0& = & 
 \rho _{,\lambda\lambda }
	 {+\frac{1}{4}\frac{\Theta}{\rho}\Big(\frac{r\rho^2\bar\sigma\sigma}{\Theta^2}\Big)_{,\lambda}}
	-\f{1}{4} \Big[ \sigma\bar F_{[\chi_{,\lambda}]}
	+\bar \sigma F_{[\chi_{,\lambda}]} \Big]
\end{eqnarray}
This gives two  hypersurface equations for the mixed derivative $\mu$ and the definition of $\mu$
\begin{eqnarray}
0 & = &   \mu _{,\lambda }
	 {+\frac{1}{4}\frac{\Theta}{\rho}\Big(\frac{r\rho^2\bar\sigma\sigma}{\Theta^2}\Big)_{,\lambda}}
	-\f{1}{4} \Big[ \sigma\bar F_{[\chi_{,\lambda}]}
	+\bar \sigma F_{[\chi_{,\lambda}]} \Big]
\\
0 & = & \rho_{,\lambda} - \mu
\end{eqnarray}

\subsection{Reduction of  additional equations for  {a} hierarchy on $\mc{B}$}
For the supplementary equation \eqref{eq:supp_r} on $\mc{B}$, we have using \eqref{eq:reformhAB,0hAB,0}
\begin{equation}
\label{eq:r_,ww}
\f{r_{,ww}}{r}	= 
- \f{1}{4}\nu\bar \nu \;\;.
\end{equation}

Evaluation of \eqref{eq:ev_r_B} on $\mc{B}$ 
 gives an evolution equation of $\Theta$ on $\mc{B}$ 
\begin{equation}
	4	\Big(r \Theta \Big)_{,w} 
	=
	 \epsilon \mathcal{R} \big|_{\mc{B}}
	  - F_{[r_{,w\lambda}]} \;\;,
\end{equation} 
where
\begin{equation}
\label{ }
F_{[r_{,w\lambda}]} =
	\mc{D}_C\Big[r^2 W^{C}_{,\lambda}\Big]
	 +\epsilon \mc{D}^F\mc{D}_F \ln r^2 
	+\f{\epsilon}{2}r^4 h_{EF}	W^E_{,\lambda}W^F_{,\lambda}	
\end{equation}
Note this equation also determines the mixed derivative $\mu$ on $\mc{B}$ through $\Theta_{,w}|_{\mc{B}} = \rho_{,\lambda}|_{\mc{B}}$ by the algebraic relation 
\begin{equation}
4r \mu 
	=\epsilon \mathcal{R} 
	-4\rho\Theta
	  + F_{[r_{,w\lambda}]} \;\;.
	  \label{eq:algbraic_mu}
	\end{equation}
The reformulation of the principal part of the evolution equation also allows us to find an evolution equation of $\sigma$ along $\mc{B}$.
Because of \eqref{eq:nu,1 = sigma,0}, we have
\begin{eqnarray}
m^Am^B(rh_{AB})_{,\lambda w} =
(r\sigma)_{,w}
\end{eqnarray} 
such that from \eqref{eq:ev_hAB} follows
\begin{equation}
\label{eq:dw_sigma_on_B}
r(r\sigma)_{,w}|_{\lambda=0}
 =  F_{[\sigma_{,w}]}(r, W^A_{,\lambda})
\end{equation}
with 
 \begin{eqnarray}
 F_{[\sigma_{,w}]}(r, W^A_{,\lambda})
 =
 \f{\epsilon r^4}{2} (m_EW^E_{,\lambda})^2
 { -\f{r^2}{2} m^E m_F \mc{D}_E W^F_{,\lambda}}
\end{eqnarray}
evaluated on $\mc{B}$.

An  evolution equation for $\mu $ along $\mc{B}$ can be found by taking the radial derivative of  \eqref{eq:r_,ww}

\begin{eqnarray}
r_{,ww\lambda }
&=&\rho_{,w\lambda }
=\mu_{,w }  \\
&=&  
	-\f{1}{4}\Theta \nu \bar \nu
	-\f{1}{4}\Big[(r \nu_{,\lambda}) \bar \nu
	+  \nu (r\bar \nu_{,\lambda})\Big]\;\;.
	\label{eq:ev_mu_bound_onB}
\end{eqnarray}
 The radial derivative $(r\nu_{,\lambda})$ on $\mc{B}$ in \eqref{eq:ev_mu_bound_onB} is found from
$$(r\sigma)_{,w} =(\rho\sigma+r\sigma_{,w}) =(\rho\sigma+r\nu_{,\lambda})$$ so that using \eqref{eq:dw_sigma_on_B} 
 \begin{eqnarray}
(r\nu_{,\lambda})|_{\lambda=0}
 &=&- \rho\sigma
	+ F_{[\sigma_{,w}]}( r, h_{AB}, W^A_{,\lambda})\;\;,
\end{eqnarray}
yields the evolution equation of $\mc{\mu}$ on $\mc{B}$ as
\begin{eqnarray}
 \mu_{,w }
&=&  
F_{[\mu_{,w}]} (r,\nu, h_{AB}, \sigma, W^A_{,\lambda})
\end{eqnarray}
where 
\begin{eqnarray}
F_{[\mu_{,w}]} & := & 
	-\f{\Theta}{4}\nu \bar \nu
	-\f{1}{4}\Big( F_{[\sigma_{,w}]} \bar \nu
	+  \nu \bar F_{[\sigma_{,w}]}\Big)
	+\f{\rho}{4}\Big( \sigma \bar \nu
	+  \nu \bar \sigma\Big)
\nonumber\\
\end{eqnarray}
vanishes in spherical symmetry.
\subsection{Summary of the metric hierarchy}
The newly derived equations for the affine, null-metric initial value problem at two intersecting null hypersurfaces split into  {three} different groups: 
\begin{enumerate}[(i)]
  \item a hierarchical set of differential equations along the null rays on the boundary $\mc{B}$ where $\lambda=0$
  \item a hierarchical set of hypersurface equations on the null hypersurfaces $\mc{N}_0$ where $w=0$
  \item  two evolution equations to propagate the initial data, i.e. the transverse traceless part of $h_{AB}$, on a given null hypersurface $\mc{N}_0$ to a null hypersurface $\mc{N}_{0+\Delta w}$
\end{enumerate}
Note, that (iii) is in difference to the traditional Bondi-Sachs approach \cite{BSscolar} in which the evolution equations evolve the transverse traceless part of the shear $h_{AB,1}$.

The set of equations on $\mc{B}$ is given by 
\begin{subequations}
 \begin{eqnarray}
h_{AB,w} & = & n_{AB} \\
\f{r_{,ww}}{r}
	&=&
	F_{[r_{,ww}]}(h_{AB}, n_{AB}) 
\\
r_{,w}& = & \rho \\
\epsilon (r^4h_{AB}W^B_{,\lambda})_{,w}
	&=& 
	F_{A[W^B_{,w\lambda}]}(r, h_{AB}, n_{AB}) 
\\
r(r\sigma)_{,w}
&=&  
F_{[\sigma_{,w}]}(r,h_{AB}, W^A_{,\lambda})\\	
4	\Big(r \Theta \Big)_{,w} 
&=&
	\epsilon \mathcal{R} 
	-F_{[r_{,w\lambda}]}(r, h_{AB}, W^A_{,\lambda})
\\
  \mu_{,w }
&=&  
	F_{[\mu_{,w}]} (r,\nu, h_{AB}, \sigma, W^A_{,\lambda})
\end{eqnarray}
\end{subequations}
the hypersurface equations on $\mc{N}_w$ are 
\begin{subequations}
  \begin{eqnarray}
  h_{AB,\lambda}&=& \sigma_{AB}
  \\
\f{1}{r}r _{,\lambda\lambda} & = &   
	F_{r_{,\lambda\lambda}}(h_{AB}, \sigma_{AB})\label{eq:HYP_hier_2}
  \\
\Theta & = &   r_{,\lambda}
\\
\epsilon\Big(r^4  h_{AB}W^B_{,\lambda}\Big)_{,\lambda}
&=&
2r^2 F_{A[W_{,\lambda\lambda}]} (r, h_{AB}, \sigma_{AB})
\\
	\Big[Y(r^2)_{,\lambda} \Big]_{,\lambda}
&=&	  \!\!\!
	  \mathcal{R}
	+F_{[Y_{,\lambda}]}(r, h_{AB}, W^A, W^A_{,\lambda})\;\;\;,
\\
- r \chi_{,\lambda}
 &=& F_{[\chi_{,\lambda}]}(r,h_{AB}, W^A,  W^A_{,\lambda}, Y)\\
\mu _{,\lambda } & = &   
	F_{[\mu_{,\lambda}]}\Big(r, \rho,\Theta,\sigma,  F_{[\chi_{,\lambda}]}\Big)
\\
\rho_{,\lambda}&=&
	\mu	
\end{eqnarray}
\end{subequations}
with 
\begin{eqnarray}
\label{ }
F_{[r_{,ww]}}(h_{AB}, n_{AB})&=&-\f{1}{4}\nu\bar \nu\\
F_{A [W^C_{,w\lambda }]}(r, h_{AB}, n_{AB}) &=&
 {2r^2\mc{D}_A\Big(\f{\rho}{r}\Big)}
-\mc{D}^B(r^2 n_{AB})
%
\nonumber\\&&\\
F_{[r_{,\lambda\lambda}]}(h_{AB}, \sigma_{AB})&=&-\f{1}{4}\sigma\bar \sigma\\
F_{A[W_{,\lambda\lambda}]}(r, h_{AB}, \sigma_{AB}) &=& 
 {-\mc{D}_A\Big(\frac{\Theta}{r}\Big)
+\frac{\mc{D}^B(r^2 \sigma_{AB})}{2r^2}}
\nonumber\\
	\\
F_{[\mu, \lambda]}	&=& -\f{1}{4} \Big[ \sigma\bar F_{[\chi_{,\lambda}]}
	+\bar \sigma F_{[\chi_{,\lambda}]} \Big]
	\nonumber\\&&
		 {+\frac{1}{4}\frac{\Theta}{\rho}\Big(\frac{r\rho^2\bar\sigma\sigma}{\Theta^2}\Big)_{,\lambda}}
\end{eqnarray}
and the evolution equations for $\lambda>0$ follow from the definitions \eqref{eq:def_YJ}, \eqref{eq:def_rho_theta_n_sigma}, \eqref{eq:decomp_n} and \eqref{eq:def_chi}
\begin{eqnarray}
  h_{AB,w} & = &\Big( \f{\chi}{r} +\f{ \rho \sigma}{\Theta}\Big)\bar m_A\bar m_B +\Big( \f{\bar \chi}{r} + \f{\rho \bar \sigma}{\Theta}\Big) m_A m_B
 \nonumber\\
\end{eqnarray}
Note, that for $w>0$, the hypersurface equation \eqref{eq:HYP_hier_2} serves as an algebraic relation because with the evolution equation  the transverse traceless part of $h_{AB}$ is evolved to later values $w>0$.

This set of equations requires the following initial-boundary data
\begin{itemize}
\item Initial data on $\Sigma$, that are functions of $x^A$, only 
\begin{equation}
\label{ }
h_{AB}\;\;,\;\;
\sigma_{AB} \;,\; r,\;\; W^A_{,\lambda} \;,\; \rho\;,\;  \Theta
\end{equation}
  \item free data on $\mc{B}$ which are functions depending on $w$ and $x^A$
	\begin{equation}
	\label{eq:free_data_B}
	  n_{AB}
	\end{equation}
  \item free data on $\mc{N}_0$ which are functions depending on $\lambda$ and $x^A$
\begin{equation}
\label{eq:free_data_N}
\sigma_{AB}
\end{equation}
\end{itemize}
The initial value for $\mu$ on a cross section $\Sigma$ is fixed at the initial time $w=0$ by \eqref{eq:algbraic_mu}. 
The variables $Y$  and $\chi$ are calculated at any time $w$ according to their definitions \eqref{eq:def_YJ} and \eqref{eq:def_chi} evaluated on $\mc{B}$ 
\begin{equation}
\label{eq:calc_Y_chi_B}
Y|_{\mc{B}} =-\f{2 \rho}{\Theta}\Big|_{\mc{B}}\;\;,\;\;
\chi|_{\mc{B}} = \Big(r\nu -\f{r\rho\sigma}{\Theta}\Big)\Big|_{\mc{B}}
\end{equation}
while the original metric function $W$ is found for ${\lambda>0}$ using \eqref{eq:def_YJ} and the known values $Y$, $\rho$ and $\Theta$.
The gauge condition $W_{,\lambda}|_{\mc{B}} = 0$, needs to be assured after the integration of the hypersurface equations. This is because  the $\lambda-$derivative of $W$ at $\mc{B}$ requires the solution for $W$ for values $\lambda>0$ and $W$ is determined by the fields  $Y,\Theta$ and $\rho$.

In particular, the functions $F_{[\cdot]}$ vanish in maximal symmetric initial-boundary data on $\Sigma$, $\mc{B}$ and $\mc{N}_0$.

\section{Examples with  spherical or hyperbolic symmetry}\label{sec:maxsym}
We assume that the metric $g_{ab}$ possesses the triad $\chi_{\mathbf{(i)}} = \chi_{\mathbf{(i)}}^{A}\partial_A$, $i\in\{1,2,3\}$,   Killing vectors, which are either generating spherical symmetry, i.e. \cite{HerreraWitten}
\begin{subequations}
\begin{eqnarray}
\label{ }
\chi^s_{\mathbf{(1)}} &=& \partial_\phi\\
\chi^s_{\mathbf{(2)}} &=& -\cos \phi \partial_\theta +\cot  \theta \sin\phi \partial_\phi\\
\chi^s_{\mathbf{(3)}} &=& -\sin \phi \partial_\theta +\cot  \theta \partial_\phi
\end{eqnarray}
\end{subequations}
or hyperbolic symmetry
\begin{subequations}
\begin{eqnarray}
\label{ }
\chi^p_{\mathbf{(1)}} &=& \partial_\phi\\
\chi^p_{\mathbf{(2)}} &=& -\cos \phi \partial_\theta +\coth  \theta \sin\phi \partial_\phi\\
\chi^p_{\mathbf{(3)}} &=& -\sin \phi \partial_\theta +\coth  \theta \partial_\phi
\end{eqnarray}
\end{subequations}
where we introduced the notation $x^A = (\theta,\phi)$, for the angular coordinates. 

As a consequence, the two space $\Sigma$ is a space of constant curvature $\mc{K}$ with 
$r^2 h_{AB}|_{\Sigma}=r^2 f_{AB}(x^C),$
where the conformal 2-metric  $f_{AB}(x^C) \in \{q_{AB},\,p_{AB}\}$ is the metric of a 2-sphere, ($q_{AB}$),  or 2-hypersphere,  ($p_{AB}$);
\begin{eqnarray}
q_{AB}dx^Adx^B & = & d\theta^2+\sin^2\theta d\phi^2\;\;, \\
p_{AB}dx^Adx^B & = & d\theta^2+\sinh^2\theta d\phi^2 \;\;,
\end{eqnarray}
where each of which gives rise to a the curvature radius $\mc{K}^{p}_{\Sigma}=-1$ and  $\mc{K}^{q}_\Sigma=1$, respectively. 
The corresponding Ricci scalars are  $\mc{R}^{p}_{\Sigma}=2\mc{K}^{p}_{\Sigma}$ and  $\mc{R}^{q}_\Sigma=2\mc{K}^{q}_{\Sigma}$. 
For notational convenience, we set  $\mc{R}_{\Sigma} = 2\mc{K}$, where $\mc{K} \in\{\mc{K}^{q	}_{\Sigma},\mc{K}^{p}_{\Sigma}\}$.
Another consequence is that the metric functions $W,\; W^A$ and $r$ are dependent of $x^A$ and that we can choose $ W^A_{,\lambda} |_\Sigma= 0$, according to the symmetry. 
Considering the free metric  data of \eqref{eq:free_data_B} and \eqref{eq:free_data_N}, the symmetry implies that two metric $h_{AB}$  does not propagate along the generators of  $\mc{B}$ and $\mc{N}_0$
\begin{eqnarray}
 \sigma_{AB}|_{\mc{N}_0} =
n_{AB}|_\mc{B} = 0 \;\;.
\end{eqnarray} 
Hence, $h_{AB, w}|_\Sigma =h_{AB, \lambda}|_\Sigma= 0$.
The initial data for the derivatives $W^A_{,\lambda}|_\Sigma$ of the shift $W^A$ are chosen to vanish. 
On the boundary null hypersurface
\begin{subequations}\label{eq:bound_spher}
 \begin{eqnarray}
h_{AB,w}|_{\mc{B}}  & = &0\label{eq:hier_B1}\;\;, \\
\f{r_{,ww}}{r}|_{\mc{B}}
	&=&
	0\;\;,
	\label{eq:hier_B2}
\\	
\rho_{\mc{B}}  &=& r_{,w}|_{\mc{B}} \label{eq:hier_B3}\;\;,\\
4	\Big(r \Theta \Big)_{,w} |_{\mc{B}}
&=&
	\epsilon \mathcal{R} |_{\mc{B}}\label{eq:hier_B4}
\;\;,\\
 \mu_{,w }|_{\mc{B}}
&=&  
	0
	\label{eq:hier_B5}\;\;,
\end{eqnarray}
\end{subequations}
the hypersurface equations on $\mc{N}_w$ are 
\begin{subequations}\label{eq:hyp_spher}
  \begin{eqnarray}
  h_{AB,\lambda}&=& 0\label{eq:hyp_S1}\;\;,
  \\
\f{1}{r}r _{,\lambda\lambda} & = &   0\label{eq:hyp_S2}\;\;,
  \\
\Theta & = &   r_{,\lambda}\label{eq:hyp_S3}\;\;,
\\
	\Big[Y(r^2)_{,\lambda} \Big]_{,\lambda}
&=&	  
	  \mathcal{R}\label{eq:hyp_S4}\;\;,
\\
\mu _{,\lambda } & = &   
	0\label{eq:hyp_S5}\;\;,\\
\rho_{,\lambda}&=&
	\mu	\label{eq:hyp_S6}\;\;,
\end{eqnarray}
\end{subequations}
and the evolution equations for $\lambda>0$
\begin{subequations}\label{eq:ev_max_sym}
\begin{eqnarray}
 h_{AB,w} & = &0
\end{eqnarray}
\end{subequations}
The initial value for $\mu$ on a cross section $\Sigma$ is fixed at the initial time $w=0$ by 
\begin{equation}
 \mu
	= \f{\epsilon\mathcal{R}}{4r} -\f{\rho\Theta}{r}
\end{equation}
following from \eqref{eq:algbraic_mu}.
The variable $Y$ can be calculated at any time $w$ on $\mc{B}$ according to \eqref{eq:calc_Y_chi_B}.
To summarise,  the (in general nontrivial) initial data to solve this system are the following (constant) scalars
\begin{subequations}
\begin{equation}
\label{ }
r_\Sigma, \;\; \rho_\Sigma,\;\; \Theta_\Sigma\;\;,
\end{equation}
as well as the (diagonal) 2-metric 
\begin{equation}
\label{ }
h_{AB}|_{\Sigma} = f_{AB}(x^A).
\end{equation}
\end{subequations}
\subsection{Integration of the  system}
\subsubsection{Solving the hierarchy on $\mc{B}$}
We begin with solving the hierarchy on $\mc{B}$. 
The vanishing of the free data $n_{AB}$ on $\mc{B}$ and integrating of \eqref{eq:hier_B1} implies
 \begin{equation}
	\label{eq:hAB_B}
	h_{AB}(w, x^A)|_{\mc{B}} = f_{AB}(x^C)\;\;,
\end{equation}
and  we also have $\mc{R}_{,w}|_{\mc{B}}=0$, i.e.  
 \begin{equation}
	\label{eq:Ric2_B}
	\mc{R}|_{\mc{B}} = \mc{R}_\Sigma\neq0\;\;.
\end{equation}
Next, integration of \eqref{eq:hier_B2} yields
  \begin{equation}
	\label{eq:r_B}
	 r_{\mc{B}} = \rho_\Sigma w + r_\Sigma \;,\;\;
	\end{equation}
and since \eqref{eq:hier_B2}  also implies $ \rho_{,w}|_{\mc{B}}=0$ we have 
	\begin{equation}
	\label{eq:rho_B}
		\rho_\mc{B}(w) = \rho_\Sigma  = const
	\end{equation}
From the integration of  \eqref{eq:hier_B3}, while using \eqref{eq:Ric2_B}, follows
  \begin{equation}
	\label{eq:Theta_B}
	\Theta_\mc{B} =\f{1}{r_\mc{B}}\Big[r_\Sigma \Theta_\Sigma+ \f{\epsilon}{4}   \mc{R}_\Sigma w   \Big]
	\end{equation}
 We now calculate  the initial value of $\mu|_\Sigma$ and integrate \eqref{eq:hier_B4}  
 to find 
 	\begin{equation}
	\label{eq:mu_B}
	\mu_\mc{B} = \mu_\Sigma = \f{ \epsilon\mc{R}_\Sigma}{4r_\Sigma}-\f{\rho_\Sigma \Theta_\Sigma}{r_\Sigma} 
	\end{equation}
	The value of $\mu_\Sigma$ also allows us to write 
	\begin{equation}
	\label{ }
	\Theta_\mc{B} = \f{1}{r_\mc{B}}\Big[\rho_\Sigma\Theta_\Sigma(w-1)+  r_\Sigma\mu_\Sigma w \Big]\;\;.
	\end{equation}
	The Ricci scalar $\mc{R}_\Sigma$ may be related to the data $r_{\Sigma}$, $\rho_\Sigma$ and  $\Theta_\Sigma$
	\begin{equation}
	\label{ }
	\mc{R}_\Sigma = - 4(\rho_\Sigma \Theta_\Sigma +r_\Sigma\mu_\Sigma)\;\;.
	\end{equation}
	At last, we calculate the boundary value of $Y$ on $\mc{B}$ using the definition \eqref{eq:def_YJ} on $\mc{B}$
	\begin{equation}
	\label{eq:Y_B}
	Y_{\mc{B}} =  2\epsilon\f{\rho_\mc{B}}{\Theta_\mc{B}}
	=\f{8 \epsilon\rho_\Sigma (\rho_\Sigma w +r_\Sigma) }{  4 r_\Sigma \Theta_\Sigma-   \mc{R}_\Sigma w}
	\end{equation}
	which is for $w=0$,
	\begin{equation}
	\label{eq:Y_B_0}
	Y (0,0) =2\epsilon\f{\rho_\Sigma}{\Theta_\Sigma}
	\end{equation}
	
The solutions \eqref{eq:hAB_B}, \eqref{eq:Ric2_B}, \eqref{eq:r_B}, \eqref{eq:Theta_B}, \eqref{eq:mu_B} and \eqref{eq:Y_B} are the boundary values necessary to integrate the hypersurface equations \eqref{eq:hyp_spher}.
\subsubsection{Solving the hierarchy on the initial data surface  {$\mc{N}_0$}}\label{sec:sol_max_sym}
We begin with integrating the data on the initial data surface  {$\mc{N}_0$} given by  $w=0$.
Integration of \eqref{eq:hyp_S1} using \eqref{eq:hAB_B} gives
\begin{equation}
\label{eq:sol_gen_hAB_w0}
h_{AB}(0,\lambda, x^A) = f_{AB}(x^A)
\end{equation}
which also implies that $\mc{R}_{,\lambda}|_{w=0}=0$, i.e. 
\begin{equation}
\label{eq:Ric2_N0}
\mc{R}(0,\lambda, x^A) = R_\Sigma = const\;\;.
\end{equation}
From \eqref{eq:hyp_S2}  and requiring $r\neq 0$ for $\lambda\ge 0$,  we have with $r_{,\lambda}|_{\mc{B}} = \Theta_\mc{B}$
  \begin{equation}
	\label{eq:sol_gen_r_w0}
          r(0,\lambda) 
          = \Theta_\Sigma \lambda+r_\Sigma\;\;.
    \end{equation}
This gives us from the definition of $\Theta$
\begin{equation}
\label{ }
\Theta(0,\lambda) = \Theta_\Sigma.
\end{equation}
Next, integrating \eqref{eq:hyp_S4}, while using \eqref{eq:Ric2_N0} and \eqref{eq:Y_B},
gives
	\begin{equation}
\label{ }
	Y
	 =	  
	\f{\mathcal{R}_\Sigma \lambda+4\epsilon r_\Sigma\rho_\Sigma }{2\Theta_\Sigma (\Theta_\Sigma \lambda+r_\Sigma) } 
	\end{equation}
	From \eqref{eq:hyp_S5}, we have
	\begin{equation}
	\label{ }
	\mu(0,\lambda) = \mu_{\Sigma}
	\end{equation}
	while \eqref{eq:hyp_S6} gives
	\begin{equation}
	\label{eq:sol_gen_rho_w0}
	\rho(0,\lambda) = \mu_\Sigma\lambda  +\rho_\Sigma
	\end{equation}
	
	Now the metric function $W$ is found readily from the definition \eqref{eq:def_YJ} using $Y, \rho$ and $\Theta$,
	\begin{eqnarray}
	\label{ }
	W(0,\lambda) &=& -\f{2\epsilon \mu_\Sigma \lambda^2}{r(0,\lambda)}
	\end{eqnarray}
	which vanishes for $\lambda=0$, as well as $W_{,\lambda}(0,0)=0$. So that the gauge condition holds on the initial null hypersurface  { $\mc{N}_0$}.

	\subsubsection{Solving the evolution equations $\mc{N}_0\rightarrow \mc{N}_w$}
	With the solutions \eqref{eq:sol_gen_hAB_w0}, we can now solve the evolution equation \eqref{eq:ev_max_sym}  to evolve the data from the initial null hypersurface  { $\mc{N}_0$ to a null hypersurface $\mc{N}_w$ }at some later `time' $w>0$. Immediate integration shows
	\begin{eqnarray}
	h_{AB}(w,\lambda, x^C) & = & f_{AB}(x^C)
	\end{eqnarray}
	This solution determines  the conformal metric $h_{AB}$ on the complete coordinate domain for the coordinate values value of $\lambda\ge 0$ and $w\ge0$ as a function of the initial data on $\Sigma$. The remaining fields must be calculated using again the hypersurface equations and the respective boundary values for $w>0$. 

	\subsubsection{Obtaining the metric on   {$\mc{N}_w$ -- Calculation} of the hypersurface equations for $w>0$ and $\lambda>0$}
From \eqref{eq:hyp_S2}  and requiring $r\neq 0$ for $\lambda\ge 0$ and $w>0$,  we have with $r_{,\lambda}|_{\mc{B}} = \Theta_\mc{B}$
  \begin{equation}
	\label{eq:sol_gen_r}
          r(w,\lambda) = \Theta_\mc{B}(w)\lambda + r_\mc{B}(w)\;\;,
    \end{equation}
which also implies using \eqref{eq:hyp_S3} 
  \begin{equation}
	\label{ }
	\Theta(w,\lambda) = \Theta_\mc{B}(w)\;\;.
  \end{equation}
  Since $h_{AB}(w,\lambda,x^A) = f_{AB}$ for all $w\ge0$ and $\lambda\ge 0$, we have 
\begin{equation}
\label{eq:Ric2_N0}
\mc{R}(w,\lambda, x^A) = R_\Sigma \;\;,
\end{equation}
which allows us to integrate \eqref{eq:hyp_S4} for $w>0$ while using \eqref{eq:Y_B}
gives
	\begin{equation}
\label{ }
	Y(w,\lambda)
	 =	  
	\f{\mathcal{R}_\Sigma \lambda+4\epsilon r_\mc{B}\rho_\Sigma }{2r\Theta_\mc{B} } \;\;.
	\end{equation}
	From \eqref{eq:hyp_S5}, we have
	\begin{equation}
	\label{ }
	\mu(w,\lambda) = \mu_{\Sigma}(w) \;\;,
	\end{equation}
	while \eqref{eq:hyp_S6} gives
	\begin{equation}
	\label{ }
	\rho(w,\lambda) = \mu_\Sigma\lambda  +\rho_\Sigma\;\;.
	\end{equation}
	
	Now the metric function $W$ for $w>0$ and $\lambda>0$ is found readily from the definition \eqref{eq:def_YJ} using $Y, \rho$ and $\Theta$,
	\begin{eqnarray}
	\label{ }
	W &=&
	\f{[ (r_\Sigma-r)\mathcal{R}_\Sigma+  4\epsilon \rho_\Sigma\Theta\Sigma  r  ] \lambda  }{2rr_\Sigma\Theta_\mc{B}} \nonumber\\
	&& + \f{ 4\epsilon \rho_\Sigma (r_\Sigma r_\mc{B}- r_\Sigma r)  }{2rr_\Sigma\Theta_\mc{B}} \;\;,
	\label{eq:sol_gen_w}
	\end{eqnarray}
	whose evaluation on $B$ gives  $W=0$, since $r(\lambda=0) = r_{\mc{B}}$.
	Calculation of $W_{,\lambda}$ on $\mc{B}$ yields
	\begin{eqnarray}
	\label{ }
	W_{,\lambda}|_{\mc{B}} &=& \f{\big(\f{\rho_\Sigma^2( \mc{R}_\Sigma-4\epsilon\rho_\Sigma \Theta_\Sigma)}{r_\Sigma} \big)}{2r^2_\mc{\mc{B}}\Theta_{\mc{B}}} w ^2 
	\nonumber\\&&
		+\f{ 2\rho_\Sigma( \mc{R}_\Sigma-4\epsilon\rho_\Sigma\Theta_\Sigma)}{2r^2_\mc{\mc{B}}\Theta_{\mc{B}}} w
	\end{eqnarray}
	whose powers of 
	 $w$ must to vanish to assure the gauge condition $W_{,\lambda}|_{\mc{B}}=0$.
	 For $r_\mc{B}\neq0$ and $\Theta_\mc{B}\neq 0$ implying 
	 \begin{equation}
	\label{ }
	r_\Sigma \neq 0,
	\end{equation}
	the result are two constraints on the initial data 
	\begin{equation}
	\label{ }
	\rho_\Sigma=0\;\;\;\mbox{or}\;\;\;
	 \mc{R}_\Sigma-4\epsilon\rho_\Sigma \Theta_\Sigma=0
	\end{equation}
	So that for $\mc{R}_\Sigma\neq0$ and $r_\Sigma\neq0$, we have the two cases 
	\begin{eqnarray}
\mbox{Case 1} & : & \rho_\Sigma \neq 0\;\; \mbox{and} \;\;\Theta_\Sigma = \f{\epsilon \mc{R}_\Sigma}{4\rho_\Sigma} \label{eq:case1}\\
\mbox{Case 2} & : &   \rho_\Sigma =0   \;\;\;\mbox{and}\;\;\; \Theta_\Sigma \;\;\;\mbox{arbitrary}\label{eq:case2}
\end{eqnarray}
	These two cases characterise two different solution of Einstein equations subject to the previous assumptions and restrictions.

	\subsubsection{Case 1}
	The case 1 solution will turn out particular simple. From \eqref{eq:case1} and \eqref{eq:mu_B} follows $\mu_{\mc{B}} = 0$, so that 
	\begin{equation}
	\label{ }
	 {\Theta_\mc{B} }= \f{\epsilon \mc{R}_0}{4 {r_\mc{B}}}(\f{ {r_\Sigma}}{ {\rho_\Sigma}}+w)\;\;\;,\;\;\;
	r_\mc{B} = \rho_\Sigma w+r_\Sigma \;\;
	\;
	 \rho(w,\lambda) = \rho_\Sigma
	\end{equation}
	Consequently, with \eqref{eq:sol_gen_r} and \eqref{eq:sol_gen_w}, we have
	\begin{eqnarray}
	r & = &\f{\epsilon \mc{R}_\Sigma}{4\rho_\Sigma} \lambda +  {\rho_\Sigma} w+r_\Sigma\;\;,\\
	W & = & 0\;\;,
	\end{eqnarray}
	which brings  the line element  into the form
	\begin{eqnarray}
	\label{eq:dn_mink}
	\lefteqn{g_{ab}dx^adx^b = 2\epsilon dwd\lambda }&&\nonumber
	\\&&+\Big[\f{\epsilon \mc{R}_\Sigma}{4\rho_\Sigma} \lambda + \rho_\Sigma w+r_\Sigma\Big]^2f_{AB}dx^Adx^B\;\;.
	\end{eqnarray}
	In deed, the coordinate transformation
\begin{equation}
\label{ }
	\lambda =- \f{2\rho_\Sigma}{\epsilon \mc{R}_\Sigma}v\;\;\;,\;\;\;
	 w= \f{(u-2 {r_\Sigma})}{2\rho_\Sigma} 
\end{equation}	casts \eqref{eq:dn_mink} into 
	\begin{equation}
	\label{ }
	g_{ab}dx^adx^b = -\f{1}{\mc{K} }  dvdu +\f{1}{4}(u-v)^2f_{AB}dx^Adx^B
	\end{equation}
	which gives the line element 
	 in `standard' flat space double coordinates, where the angular coordinates $x^A$ either parameterises a 2-sphere ($f_{AB} = q_{AB}, \mc{K}=1$) or 2-hypersphere ($f_{AB} = p_{AB}, \mc{K}=-1$).
	 The Weyl scalar \eqref{psi2} obviously vanishes for the metric of \eqref{eq:dn_mink}.
	\subsubsection{Case 2}
	In case 2, where 
	 $r_\Sigma \neq 0$, 
	$\rho_\Sigma=0$,  and  $ \Theta_\Sigma$ being  arbitrary, the boundary values $\mu_\mc{B}$, $r_\mc{B}$ and $\Theta_\mc{B}$ are 
	\begin{eqnarray}
	\label{ }
	\mu_\mc{B}&=&\f{\epsilon { \mc{R}_\Sigma}}{4 {r_\Sigma}}\;\;\;,\;\;\\
	r_\mc{B} &=& r_\Sigma\;\;,\;\;\\
	\Theta_\mc{B} &=& \Theta_\Sigma + \f{\epsilon \mc{R}_\Sigma}{4r_\Sigma}w\;\;.
	\end{eqnarray}
	They give the metric functions $r$ and $W$ as
	\begin{eqnarray}
	r &=& r_\Sigma +\Big[\Theta_\Sigma+\mu_\Sigma w\big]\lambda\;\;,\\
	Y 
	&=& 
	2\epsilon \mu_\Sigma\Big(\f{ \lambda r_\Sigma  }{r\Theta_\mc{B}} \Big)\;\;,\\
	W &=& -\Big(\f{ \mathcal{R}_\Sigma}{2r_\Sigma}\Big)\f{\lambda^2   }{r} \;\;.
	\end{eqnarray}

	Calculation of the Weyl scalar $\Psi_2$ shows
	\begin{equation}
	\label{ }
	\Psi_2=\f{\epsilon r_\Sigma^2\mu_\Sigma  }{r^3 }= \f{ r_\Sigma \mathcal{R}_\Sigma   }{4r^3 }	 = \f{r_\Sigma\mc{K}}{2r^3}\;\;.
	\end{equation}
	Since $\lim_{R\rightarrow\infty } R^3\Psi_2 = M$ for a Schwarzschild spacetime\footnote{Here we use in in the contrary to NP a positive signature, in NP there is $\Psi_{2.NP}=-M/R^3$. }, where $R$ is the standard Schwarzschild area distance  coordinate and $\Psi_2 = M/R^3$, we set
	\begin{equation}
	\label{ }
	 r_\Sigma = 2M\mc{K}\;\;,\;\;
	 { \Theta_\Sigma=0}
	\end{equation}
	 then the line element for the case 2 solution is 
	\begin{eqnarray}
	\label{eq:dn_BH}
		g^{(BH)}_{ab}dx^adx^b &:=&  \Big(\f{ \mc{K}}{2M}\Big)\f{\lambda^2   }{r} dw^2 + 2\epsilon dwd\lambda \nonumber\\
		&&+ r^2f_{AB}dx^Adx^B\\
		\mbox{with}&& r := \mc{K}\Big(2M +\f{\epsilon w\lambda }{4M}\Big)\;\;,
	\end{eqnarray}
	where $\mc{K}^2=1$ was used.
	Note, the line element \eqref{eq:dn_BH} is in fact independent of $\mc{K}$.
	In spherical symmetry, where $f_{AB} = q_{AB}$, the line element \eqref{eq:dn_BH} is the Israel-Schwarzschild solution \cite{Israel1}. 
	This somewhat unknown solution was derived by Israel by means of coordinate transformations form the standard Schwarzschild metric in polar coordinates 
	\footnote{The standard time-symmetric Schwarzschild form with coordinates $(T,R,X^A)$ is obtained from \eqref{eq:dn_BH} by the coordinate transformation $w =U\sqrt{\f{R}{2M}-1}$, $\lambda =  \epsilon (4 M)U^{-1}\sqrt{2M(R-2M)}$ and $X^A = x^A$ with ${U=\exp(\f{\epsilon T +R}{4M})}$. }. 
	It represents a global covering of the Schwarzschild space time like the Kruskal-Szekers solution, but has the advantage of being expressed in terms of simple rational functions with respect to the coordinates rather than an implicit relation between them, as in Kruskal-Szekers solution. 
	The horizon $\mc{H}$ of the Schwarzschild black hole is given by $\lambda=0$, i.e. $\mc{H} = \mc{B}$. 
	While the physical singularity $r=0$ is given by $ w\lambda = -8\epsilon M^2 $.
	The causal structure and Penrose-Carter diagrams of regarding the Israel--Schwarzschild solution and some of its extensions can be found in the literature \cite{Israel1,Israel2,isreal_sol_all,PajerskiNewman}. 
	The solution had also been employed to find  initial data for numerical relativity \cite{CompleteNull}, i.e. to find the null data for black hole collision. 	

	In hyperbolic symmetry, where $f_{AB} = p_{AB}$, the solution represents (after change of signature of $g_{ab}$) the interior solution $(r<2M)$ of a black hole, when the Israel coordinate transformations \cite{Israel1,PajerskiNewman} are 	applied to a solution of the Schwarzschild spacetime  recently proposed by Herrera and Witten \cite{HerreraWitten}. 
	
	It is interesting to determine a timelike Killing vector $\chi_{\mathbf{s}}$ for \eqref{eq:dn_BH} in the limit $r\rightarrow\infty$ by considering the Killing equation for \eqref{eq:dn_BH}.
	Requiring $ g^{(BH)}_{ab}\chi_{\mathbf{s}}^a\chi_{\mathbf{s}}^b = -1$ for $r\rightarrow \infty$, we find from  Killing equations that
	\begin{equation}
	\label{eq:static_Killing}
	\chi_{\mathbf{s}} = \f{1}{2M}(w\partial_w-\lambda \partial_\lambda )\;\;,
	\end{equation}
	which is null on the horizon $\mc{H}$ . 
	In particular, using a standard expression \cite{Wald}to calculate the surface gravity $\kappa$ on the horizon
	\begin{equation}
	\label{ }
	\kappa^2 =-\f{1}{2}g^{ac}_{(BH)}g^{bd}_{(BH)} (\nabla_a \chi_{\mathbf{s}b} ) (\nabla_c \chi_{\mathbf{s}d} )|_{\lambda=0}
	\end{equation}
	we find the common result $\kappa = (4M)^{-1}$. 
	A local power series expansion of \eqref{eq:dn_BH} shows that, alternatively, in this coordinate chart $x^a$
	\begin{equation}
	\label{ }
	\kappa^2  = \f{1}{4} g^{(BH)}_{ww,\lambda\lambda} |_\mc{B}\;\;.
	\end{equation}

      It is of technical interest to look at the asymptotic behaviour of the solution \eqref{eq:dn_BH} for $\lambda\rightarrow\infty$ as well as $u\rightarrow\infty$ in respect of the Penrose conformal compactification \cite{ConfInfPenrose}. 
      In the conformal compactification the null boundary surface $\mc{I}$ in the limit $r\rightarrow \infty$ is the `edge' $\mc{I}:=\partial\mc{M}$ of the manifold $ \mc{M}$ and the union $\mc{I}\cup\mc{M}$ defines a `bigger' unphysical manifold $\hat{\mc{M}}$ with a conformal metric $\hat  g_{ab} = \Omega^2 g_{ab}$,  where the conformal factor $\Omega $ obeys $\Omega|_{\mc{I}} = 0$ and $\nabla_a\Omega|_{\mc{I}}\neq 0$. 
      
      We observe that both $g_{ww}$ and $g_{AB}$ are infinite  if $\lambda\rightarrow \infty $, while only  $g_{AB}$ is infinite if $u\rightarrow\infty$. 
      For a conformal compactification,  we could in principle proceed in the in two different ways. The first approach uses  a transformation to a Bondi-Sachs metric and then standard techniques to compactify the resulting metric as e.g. outlined in \cite{BSscolar} and implemented in numerical codes (see \cite{JeffLRR} for a review). The second approach is to straightforwardly compactify the coordinates $w$ and $\lambda$ and define another suitable conformal factor. 
      
      Hereafter, we illustrate both methods for the metric \eqref{eq:dn_BH} where  $f_{AB}  $ is the unit sphere $q_{AB}$, therefore $\mc{K}=1$, and the $w-$coordinate corresponds to the retarded time $u$, so that $\epsilon=-1$. Hence the metric has the non-zero components
      \begin{eqnarray}
g_{uu} &=&\f{\lambda^2}{2Mr(u,\lambda)}
\;\;,\;\;g_{u\lambda}  =  -1 \;\;,\;\;
g_{AB}  = r^2(u,\lambda)q_{AB}\nonumber\\
&&\label{eq:israel}\\\
\mbox{with}&& r(u,\lambda) = 2M -\f{u\lambda}{4M}\;\;,\label{areadist}
\end{eqnarray}
      or alternatively expressed directly in terms of the coordinates
      \begin{subequations}\label{eq:is_explicite}
      \begin{eqnarray}
g_{uu} &=&\f{2\lambda^2}{8M^2-u\lambda}\\
g_{u\lambda} & = & -1 \\
g_{AB} & = & \Big(\f{1}{4M}\Big)^2\Big(8M^2 - u\lambda\Big)^2q_{AB}
\end{eqnarray}
      \end{subequations}
	We wish to transform  the radial coordinate, the   affine parameter $\lambda$, to an inverse area distance coordinate $\ell$ \cite{tam,BSscolar}. 
	Then, this inverse area distance is a preferred conformal factor $\Omega:=\ell$ to perform the Penrose compactification and find the conformal metric $\hat g_{ab} = \ell^2g_{ab}$.
	We observe that the surface area of the 2-surfaces $u=const$ and $\lambda=const$ in \eqref{eq:israel} is $4\pi r^2(u,\lambda)$.
	Therefore the inverse of conformal factor $r$ is the necessary inverse area distance coordinate $\ell := 1/r$. 
	Due to the explicitly known relation \eqref{areadist}, we can invert $\ell =1/r(u,\lambda)$ one-to-one to find a $\lambda(u, \ell)$ and the respective coordinate change from $g_{ab}(u,\lambda, x^A)\rightarrow g_{ab}(u, \ell, x^A)$ using 
	\begin{equation}
	\label{ }
	d\lambda = -\f{\lambda du}{u}+\f{4M d\ell}{u\ell^2}
	\end{equation}
	so that 
	 \begin{eqnarray}\label{eq:time_dependent_scri}
	g_{uu} &=&\f{\lambda^2 \ell }{2M} - \f{2\lambda }{u}
	\;\;,\;\;g_{u\ell}  =  -\f{4M}{u\ell^2} \;\;,\;\;
	g_{AB}  = \f{q_{AB}}{\ell^2}\nonumber\\
\end{eqnarray}
	This metric has the undesirable feature that $g_{u\ell}$ is an explicit function of $u$.
	This ambiguity can be removed with a remapping of the $u$ coordinate  $u = u(U)$.
	The requirement on $g_{U\ell} = \ell^{-2}$ gives
	\begin{equation}
	\label{ }
	\f{d u}{d U } = -\f{1}{\ell^2 \lambda_{,\ell}} = -\f{u}{4M}
	\end{equation}
	which implies
	\begin{equation}
	\label{ }
	u(U) = e^{-\f{U}{4M}}\;\;.
	\end{equation}
	Therefore the combined coordinate transformation $(u,\lambda,x^A)\rightarrow(U,\ell,x^A)$ maps the line element of \eqref{eq:israel} to
	\begin{eqnarray}
	ds^2& = & \f{1}{\ell^2}\Bigg\{(-\ell^2+2M\ell^3)dU^2+2dUd\ell 
	+ q_{AB}dx^Adx^B\Bigg\} \nonumber\\\label{eq:phys_met_compact}
	\end{eqnarray}
	The  conformal line element $d\hat s^2 = \Omega^2ds^2$ corresponding to \eqref{eq:phys_met_compact} is with $\Omega:=\ell$
	\begin{eqnarray}
	d\hat s^2& = &  (-\ell^2+2M\ell^3)dU^2+2dUd\ell 
	+ q_{AB}dx^Adx^B\;\;, \nonumber\\\label{eq:EF_conf}
	\end{eqnarray}
	which is the `standard' conformal representation of the Schwarzschild metric \cite{ConfInfPenrose}. 
	In particular, $d\hat s^2$ is the conformal Minkowski metric for $\ell=0$. This metric can now be used to analyse the neighbourhood of $\ell=0$ corresponding to the limit $r\rightarrow\infty$. For example, we find the Weyl scalar $\Psi_2$ for an adapted null tetrad of \eqref{eq:EF_conf} as being simply given\footnote{Note, for a conformal transformation $\hat g_{ab}=\Omega^2g_{ab}$, the Weyl scalar $\Psi_2$ behaves as $\hat\Psi_2 =\Omega^2\Psi_2$ see e.g. \cite{StewardBondiMass}} by $\hat \Psi_2 = M\ell^3$. 
	
	Another procedure to extract the physical information in the asymptotic regions may be found by a 
	second approach. Since $r\rightarrow \infty$ for both $u\rightarrow	\infty $ and $\lambda\rightarrow	\infty $, the strategy is to compactify both coordinates $u$ and $\lambda$, and define new coordinates $U = 2/u$ and $\Lambda = -2/\lambda$. Then the metric \eqref{eq:is_explicite} becomes  {under} the transformation $(u,\lambda, x^A)\rightarrow(U,\Lambda, x^A)$
	      \begin{eqnarray}
		g_{UU} &=&\f{8}{U^3\Lambda(2M^2U\Lambda+1)}\\
		g_{U\Lambda} & = & \f{4}{U^2\Lambda^2} \\
		g_{AB} & = & \Big(\f{2M^2U\Lambda +1}{4MU\Lambda}\Big)^2 q_{AB}
	\end{eqnarray}
	To find a conformal metric $\hat g_{ab} = \Omega^2 g_{ab}$, we make the {\it naive} choice of $\Omega = U\Lambda$ for the conformal factor 
	The metric at $\Lambda=0$ is {\it not } the Minkowski metric, but
	\begin{eqnarray}
	\hat g_{ab}dx^adx^b|_{\Lambda=0}& = & 4 dUd\Lambda +\f{q_{AB}dx^Adx^B}{M^2}\;\;,\label{eq:2nd_conf}
	\end{eqnarray}
	however the physical information encoded in the conformal curvature tensor by means of leading order of $\hat \Psi_2$ is simply given by 
	$$\hat \Psi_2(U,\Lambda) = M^4 (U\Lambda)^3+O(\Lambda^4)\;\;. $$

	Regarding numerical simulations the two  presented approaches to extract the asymptotic quantities, have their advantages and disadvantages: The first approach, the Bondi-Sachs approach, is rather straightforward but it has the `disadvantage' that the coordinate transformation from the affine parameter is general dependent dependent on $w$ and $x^A$ (The $w$ dependence is illustrated, e.g. in \eqref{eq:time_dependent_scri}). This yields to nontrivial shifts $U^A$ with respect to the generators of the null hypersurface $\mc{I}$, for which however exist recipes \cite{news,Handmer, BSscolar} to resolve this issue. In the second approach, the trivial coordinate compactification of $u$ and $\lambda$ together with the naive choice of the product of the new coordinates as conformal factor is simple to implement numerically. However, the resulting asymptotic structure of the metric at $\mc{I}$ is not straightforward because a Minkowski metric does not arise naturally in this procedure. 
	In particular, asymptotic symmetries like the BMS symmetries seem to be difficult to single out.
	Nevertheless, applying the second procedure to  Israel black hole solution \eqref{eq:israel} show in \eqref{eq:2nd_conf} that the inverse of the  black hole mass squared may already be read off at leading order of the conformal 2-metric $\hat h_{AB}$. 
	The latter may be beneficial in numerical work, as the extraction of the mass in \eqref{eq:EF_conf} requires to calculate a third order radial derivative to pick out the $O(\ell^3)$ coefficient, while in \eqref{eq:2nd_conf} the mass parameter may be read of from the metric at $\Lambda=0$.

\section{Discussion}
An initial-boundary formulation for the affine, null-metric formulations was derived for a family of null hypersurfaces that is attached to a null boundary surface. 
The formulation is well suited for the  study of  spacetimes of isolated black holes (or white holes) for which the null boundary surface is a horizon. 
The free initial data for this formulation are the set of three scalar  {functions}, one 2-vector field and two 2-tensor fields  { on a} common intersection of the null hypersurfaces, as well as one 2-tensor  {field} on each of the two null hypersurfaces having this common intersection. 
The latter 2-tensor fields are geometrically the shear of the each of the respective null hypersurfaces. 
The differential equations that determine the boundary values as well as the evolution of the initial data are  ordered sets of differential equations for both the boundary evolution and the evolution of the initial data off the null boundary surface.
This hierarchical structure is similar like the one employed in current numerical codes using the Bondi-Sachs formalism. 
However the advantage of the presented formalism to the Bondi-Sachs formalism is that the latter includes (due to the choice of coordinates) a metric variable, the expansion rate of the null hypersurfaces, which becomes singular in the presence of a horizon and thus yields inferior numerical accuracy \cite{Husa_retarded}. 
In the affine, null formulation, such behaviour is prevailed, because  affine  {parameters} are used to parameterise two coordinate directions.

The formalism has manifold possibilities for application both on the numerical as well as analytical side.
Prior to numerical work it may be interesting to look for exact linear solutions \cite{master}, that can used as test best solution for a nonlinear numerical code. 
Numerical application for the formalism are, for example, the characteristic evolution of isolated, perturbed black holes formed after a  merger of compact objects (e.g. binary black hole merger, neutron star merger) with the corresponding gravitational  wave extraction at null infinity. 
To to so, the boundary surface would be the horizon of the newly formed black hole and a compactified version of the formulation should be used like those presented for the Israel black hole solution in Sec. ~\ref{sec:sol_max_sym}. 
Since a null boundary surface is essential for the initial-boundary value formulation presented here, the formalism is particular well-adapted to investigate (at least from the classical level) some features the soft-hair proposal of Hawking and collaborators \cite{HPS}. 
Essentially the question how BMS-type supertranslations on the black hole horizon  \cite{Donnay1,Donnay2}, relate to BMS supertranslation\cite{Bondi,SachsBMS} or their extensions\cite{ext} at null infinity can be well studied.
As such current `hot' questions on how these horizon-supertranslations related to gravitational wave memory \cite{mem} and angular momenta \cite{ang} at null infinity may be addressed.

\section{Acknowledgments}
The idea of this work was initiated while I attended a particle physics conference with C. Malone of University of Cambridge, whose support during that time and afterwards is greatly appreciated. 
I am happy  to thank Joan Camps and Miguel Pino for discussions at different stages of this project, some of which had been very motivational. 
It is a particular  {pleasure for me} to thank  Jeff Winicour for constant support and teaching me the various facets of the Bondi-Sachs  formulation of General Relativity.  
I  am grateful to the authors of \cite{curvature_alpha}, in particular K. Prabhu, for communicating their results and comments on the manuscript. Comments from G. Esposito and F. Alessio are also well appreciated. The  {inclusion} of App.~\ref{app:gen_null}-\ref{sec_confRic} was motivated after some email exchanges with D. Nichols.

\begin{appendix}


\section{Derivation of Ricci tensor components for a general metric at a null hypersurface}\label{app:gen_null}
Let $(\mc{M},g)$ be a  four dimensional spacetime with a smooth metric $g$. Let $\mc{N}$ be a a family of null hypersurfaces in $\mc{M}$ labeled by the scalar function $x^0$. We further assume that $\mc{N}$ is free of caustics or crossovers of the rays generating $\mc{N}$ meaning the expansion rates of the null generators is everywhere nonzero on $\mc{N}$ (see e.g. \cite{caustic} for a discussion on caustics).  

A hypersurface $\mc{N}$ represented by $x^0=const$ is  a null hypersurface if the norm of the gradient $k_a = \nabla_a x^0$ vanishes, i.e.
\begin{equation}
\label{eikonal}
g^{ab}k_a k_b=g^{ab}(\nabla _a x^0)(\nabla_b x^0)=0
\end{equation}
Then by metric duality, the  vectors $k^a=g^{ab}\nabla_b x^0$ are normal to the surfaces, hence $k^a$ is self-orthogonal. 
The null curves generated by $k^a$  {in a} given null hypersurface of  {$\mc{N}_{x^0}$} are called rays. 
Take $x^0$ to be the first component of the coordinate vector $x^a$, then we have  $k^a = g^{a0}$ by \eqref{eikonal}. 
Choose $x^A,\;\;A=2,3$ as two additional parameters that are constant along a ray with  tangent vector $k^a$, that is the Lie transport of $x^A$ along $k^a$  vanishes
\begin{equation}
\label{const_angles}
\mc{L}_k x^A = k^a\nabla_a x^A = g^{ab}(\nabla_a x^0)(\nabla_b x^A)=0
\end{equation}
where $\mc{L}_k$ denotes the Lie derivative along $k^a$. We  take $x^A$ as coordinate scalars. 
The null and constancy conditions \eqref{eikonal} and \eqref{const_angles}, respectively,  imply
\begin{equation}
\label{ }
g^{00} = g^{0A} = 0\;\;.
\end{equation} 
The nonzero contravariant components $g^{ab}$ of the metric tensor  are 
\begin{equation}
\label{gab_up_matrix}
g^{01}\;\;,\;\;g^{11}\;\;,\;\; g^{1A}\;\;,\;\; g^{AB}
\end{equation}
so that the determinant of the corresponding matrix is 
\begin{equation}
\label{ }
\det(g^{ab}) = -(g^{01})^2\det(g^{AB})\;\;.
\end{equation}
As the covariant  components of the metric are inverse to the contravariant components and the calculation of the inverse of the matrix $(g^{ab})$ represented by \eqref{gab_up_matrix} requires a finite determinant we demand  
\begin{equation}
\label{metric_req}
g^{01}\neq 0\;\;\mbox{and}\;\;
\det(g^{AB})\neq 0
\end{equation}
and no further restriction shall be made as this stage. 

For the calculation of the covariant components $g_{ab}$ we set
\begin{equation}
\label{ }
g^{01}:=\epsilon|g^{01}|\;\;,\;\;
g^{11}:=-Vg^{01}\;\;,\;\;
g^{1A}:=U^Ag^{01}\;\;,\;\;
\end{equation}
with $\epsilon=\pm1$.
Using the completeness  relation $g^{ac}g_{cb}=\kron{a}{b}$ yields
\begin{subequations}
\begin{eqnarray}
1&=&\kron{0}{0} 
\;\;\Rightarrow\;\; g_{01}=\f{1}{g^{01}}   \\
0&=&\kron{0}{1} 
\;\;\Rightarrow\;\; g_{11}=0 \\
0&=&\kron{0}{A}  
\;\;\Rightarrow\;\; g_{1A}=0 \\
\kron{A}{B}  &=&  g^{Aa}g_{aB} 
\;\;\Rightarrow\;\; g^{AC}g_{CB}=\kron{A}{B}\\
0&=&\kron{1}{A} 
\;\;\Rightarrow\;\; g_{0A}=-U_A \\
0&=&\kron{1}{0}  
\;\;\Rightarrow\;\; g_{00}  =  Vg_{01}+ U_AU^A
\end{eqnarray}
\end{subequations}
where $U_A = g_{AB}U^A$. 
The corresponding line element is 
\begin{eqnarray}
g_{ab}dx^adx^b & = &\Big( Vg_{01} + U_AU^A\Big)(dx^0)^2+2g_{01}dx^0dx^1  \nonumber \\
 & & -2U_A dx^0 dx^A + g_{AB}dx^Adx^B\;\;.\label{eq:gen_null_metric}
\end{eqnarray}
For the determinant of the metric $g_{ab}$, we find
\begin{equation}
\label{ }
g:= \det(g_{ab}) = -(g_{01})^2 f \qquad
\end{equation}
with $$ f:=\det(g_{AB})=\f{1}{\det(g^{AB})}\;\;.$$
where the derivative
\begin{equation}
\label{ }
g^{AB}g_{AB,a} =(\ln f)_{,a}\;\;.
\end{equation}
An immediate consequence of the  coordinate conditions $g^{00}=g^{0A}=g_{11}=g_{1A}=0$ is
that the following Christoffel symbols of vanish
\begin{equation}
\label{ }
\chr{0}{1a} \;\;,\;\;
\chr{A}{11}\;\;.
\end{equation}
The remaining Christoffel symbols for \eqref{eq:gen_null_metric} are listed in App.~\ref{sec:Christ}, in what follows,  we only need
\begin{equation}
\label{christ_constraint}
g^{ab}\chr{0}{ab} =- g^{01}(\ln \sqrt{f})_{,1} \neq 0\;\;,
\end{equation}
whose non-vanishing character is required hereafter.
The twice contracted Bianchi identities are $B_a = 0$, where 
\begin{equation}
\label{ }
B_b:=g^{ab}\nabla_a\Big[R_{bc} - \f{1}{2}g_{bc} (g^{ef}R_{ef})\Big]
\end{equation}
with  $R_{ab}$ being the Ricci tensor. On a given null hypersurface $\mc{N}(x^0)$, where $x^0=const$, we assume the following six main equations 
\begin{subequations}\label{eq:main}
\begin{eqnarray}
0&=&R_{11}\Big|_\mc{N}  =  R_{1A}\Big|_\mc{N}  = g^{AB}R_{AB}\Big|_\mc{N} \label{eq:hyp}\\
0&=&\Big[R_{AB} - \f{1}{2}g_{AB}(g^{CD}R_{CD})\Big]\Big|_\mc{N}\label{eq:mainRAB}
\end{eqnarray}
\end{subequations}
hold in the following. Then from $B_1=0$ follows
\begin{equation}
\label{ }
0=-g^{ab}\chr{c}{ab}R_{c 1} = g^{01}(\ln \sqrt{f})_{,1} R_{01}\;\;.
\end{equation}
Since $ g^{01} (\ln \sqrt{f})_{,1}\neq 0$ as of \eqref{christ_constraint}, 
we have that 
\begin{equation}
\label{eq:triv_R01}
R_{01}=0
\end{equation} 
is trivially fulfilled. 
The equation $R_{01}=0$ is thus an algebraic consequence if  the main equations hold, it it called the trivial equation.
Now consider $B_A=0$ giving
\begin{equation}
\label{ }
0 = g^{01}\Big(R_{0A,1}-R_{01,A}\Big) -g^{ab} \chr{c}{ab}R_{eA}\;\;,
\end{equation}
employing $R_{01}=0$ and calculation of the Christoffel symbol yields
\begin{equation}
\label{ }
0 = \f{g^{01}}{\sqrt{f}}\Big(\sqrt{f} R_{0A}\Big)_{,1}\;\;.
\end{equation}
This equation shows that if $R_{0A}=0$ for one cut of $\mc{N}$ at an arbitrary  value for $x^1$, there is $R_{0A}=0$ for all other values of $x^1$ on $\mc{N}$. Without loss of generality, we thus require 
\begin{equation}
\label{eq:supp_R0A}
R_{0A}\big|_{x^1=0} =0\;\;.
\end{equation}
Proceeding next with $B_0=0$, we find under the assumption of \eqref{eq:main}, \eqref{eq:triv_R01} and \eqref{eq:supp_R0A}
\begin{equation}
\label{ }
0 = \f{g^{01}}{\sqrt{f}}\Big(\sqrt{f} R_{00}\Big)_{,1}\;\;.
\end{equation}
Here, the same arguments as for $R_{0A}$ apply; $R_{00}=0$ holds everywhere on $\mc{N}$ provided it holds on one arbitrary cut $x^1=const$. Henceforth, we require
\begin{equation}
\label{eq:supp_R00}
R_{00}\big|_{x^1=0} =0\;\;.
\end{equation}
The three equations \eqref{eq:supp_R00} and \eqref{eq:supp_R0A} are called supplementary equations. This terminology of grouping the Einstein equations, was introduced by Bondi and Sachs in their pioneering articles on the characteristic formulation of General Relativity \cite{Bondi,Sachs,Sachs_civp,tam,BSscolar}.
Regarding the calculation of the Ricci tensor to form the relevant field equations, we only need to calculate the main equations for all values of $x^a $ and the supplementary equations at the particular value $x^1=0$.

Instead of calculating explicitly the traceless part of $R_{AB}$, it is useful to introduce a complex dyad $t_A$, which obeys
\begin{equation}
\label{eq:def_tA}
g_{AB} =\f{1}{2}( t_A\bar t_B + t_B\bar t_A )\;\;,\;\;
t_A\bar t^A-1 = t_At^A =0\;\;.
\end{equation}
Then $R_{AB}$ can be expressed  as
\begin{eqnarray}
\label{ }
R_{AB} &=& (t^Et^F R_{EF})\bar t_A\bar t_B + (\bar t^E\bar t^F R_{EF}) t_At_B 
\nonumber\\&&+\f{g_{AB}}{2}(g^{EF} R_{EF})\;\;,
\end{eqnarray}
so that \eqref{eq:mainRAB} can be represented by either of the two equations
\begin{equation}
\label{ }
0=t^Et^F R_{EF}\;\;,\;\;
0=\bar t^E\bar t^F R_{EF}\;\;.
\end{equation}
The Ricci tensor components for \eqref{eq:hyp}, \eqref{eq:supp_R00} \eqref{eq:supp_R0A} and ${t^Et^F R_{EF}=0}$ are calculated in App.~\ref{sec_genRic}.
App.~\ref{sec_confRic}, then specifies these components for the conformally rescaled metric 
\begin{equation}
\label{eq:conf_gAB}
g_{AB} = r^2 h_{AB}\;\;,\;\;
h_{AB} = \f{1}{2}(m_{A}\bar m_{B}+\bar m_{A} m_{B})
\end{equation}
where we set $\det(h_{AB}) = h(x^A)$ and ${t_A = r m_A}$.

In particular, the Ricci tensor components for the main equation w.r.t the metric \eqref{eq:gen_null_metric} are given in \eqref{eq:R11_gen},\eqref{eq:R1A_gen}, \eqref{eq:trRAB_gen} and \eqref{eq:ttRAB_gen} while those for the  supplementary eqautions are \eqref{eq:R00_gen} and \eqref{eq:R0A_gen}.

These main equations can be easily specified for any other parametrisation of $g_{ab}$. 

Hereafter, we consider the two most `prominent' choices -- the Bondi-Sachs and an affine null metric. 
For this purpose we investigate the relation between $g^{01}$ and the null vector $k^a$.
In the current choice of coordinates, the tangent vector, $k^a=g^{ab}\nabla_b x^0$, takes the  form  
\begin{equation}
\label{ }
k^a = (0,g^{01}, 0, 0)\;\;
\end{equation}
showing that the requirement 
$g^{01}\neq 0$
assures that the tangent vector of the generators of the null hypersurfaces $x^0=const$ never vanishes and is everywhere non-zero on $x^0=const$. 
The expansion rate of the null vector $k^a$ is 
\begin{equation}
\label{ }
\theta(k) := \nabla_a k^a  = \f{g^{01}}{\sqrt{f}}(\sqrt{f})_{,1}
\end{equation}
Hence for the initial requirement,  for a caustic-free null hypersurfaces $\mc{N}$, we necessarily need  $g^{01}\neq 0$, $ {\sqrt{f} \neq 0}$  and $(\sqrt{f})_{,1}$ on $\mc{N}$. 

Pick a null ray with $x^0=const$ and $x^A=const$, and let $\lambda$ be an affine parameter along this null ray, then the ratio $dx^1/d\lambda$ is a function of the coordinates $x^a$
\begin{equation}
\label{ }
\f{d x^1}{d\lambda} = g^{01}, 
\end{equation}
 meaning 
$g^{01}$ measures the change of the  ray parameter  $x^1$ relative to change of an affine parameter $\lambda$ \cite{Ellis85}.
In the definition $g^{01}:=\epsilon |g^{01}|$, the meaning of $\epsilon$ may be understood as follows: Let $u^a$ be a future pointing timelike unit vector field parameterised with the proper time $x^0$. Then , the normalisation $k_au^a = \epsilon$, while keeping  $x^1=const$ and $x^A=const$, implies that $k^a$ is future pointing, if $\epsilon=-1$,  and past pointing, if $\epsilon=1$. Consequently, $x^0=const$ are outgoing null hypersurfaces, if $\epsilon=-1$, and  ingoing null hypersurfaces, if $\epsilon=1$.

For the affine null metric of Sec.~\ref{sec:dn_system}, we set $x^1 = \lambda$ such that   the coordinates are $x^a = (w,\lambda, x^A)$ and we make the transformations
\begin{equation}
\label{eq:spec_anm}
g_{w\lambda} = \epsilon\;\;,\;\;
W = -\epsilon V\;\;,\;\;
 {W^A = U^A\;\;.??}
\end{equation}
The corresponding line element is \eqref{eq:dn_metric}, i.e.
\begin{eqnarray}
\lefteqn{g_{ab}dx^adx^b = -W dw^2+2\epsilon dwd\lambda }\nonumber \\
 & &+r^2h_{AB} (dx^A-W^Adw)(dx^B-W^Bdw)\;\;.
\end{eqnarray}

The Bondi-Sachs metric has an area distance $r_A$ as coordinate along the rays, which is defined via  
 the expansion rate $\theta(k)$ of the null rays $k^a$.
The expansion rate $\theta(k)$ indicates the relative change of an area $\delta A$ of  cross section of a bundle of rays as measured by two neighbouring observers with an affine parameter distance $d\lambda$. 
This defines an area distance $r_A$ as \cite{Jordan}
\begin{equation}
\label{def_r}
k^a \nabla _a \ln r_A :=  -\f{1}{2} \theta(k) 
\;\;.
\end{equation}
For a  conformal decomposition $g_{AB} = r^2h_{AB}$, we have 
\begin{equation}
\label{theta_k}
\theta(k) = g^{01} \{(\ln r^2)_{,1} + [\ln\sqrt{\det{h}_{AB}}]_{,1}\}\;\;.
\end{equation}
so that 
if $(\sqrt{\det{h}_{AB}})_{,1} = 0$, \eqref{def_r}  and \eqref{theta_k} imply
\begin{equation}
\label{ }
\partial_1 \ln r^2_A = \partial_1 \ln r^2 \;\;.
\end{equation}
Hence, the area distance $r_A$ is basically the conformal factor $r$ of the conformal decomposition of the  two metric $g_{AB}$ . 

Choosing $\sqrt{\det{h}_{AB}} = h(x^A)$ and  $x^1 = r_A$ as coordinate along the rays gives the traditional Bondi-Sachs metric \cite{BSscolar}  with 
 coordinates $x^a = (u,r:=r_A,x^A)$ and  $\epsilon = -1$,  $g_{ur} = -e^{2\beta}$, $V\rightarrow V/r$ and $g_{AB} = r^2 h_{AB}$, while $r_{,u}=r_{,A}=0$. The line element is
\begin{eqnarray}
g_{ab}dx^adx^b & = &\Big(-\f{V}{r}e^{2\beta}+r^2h_{AB}U^AU^B\Big)du^2-2e^{2\beta}dudr \nonumber \\
 & &-2r^2h_{AB} U^Adudx^A +  r^2h_{AB}dx^Adx^B\;\;.\label{BSmetric}
\end{eqnarray}

If $g_{AB}=r^2h_{AB}$ the corresponding components to relevant Ricci tensor components  are \eqref{eq:R11_h},\eqref{eq:R1A_h}, \eqref{eq:trRAB_h} and \eqref{eq:ttRAB_h} for the main equations, and  \eqref{eq:R00_h} and \eqref{eq:R0A_h} for the supplementary equations. 

After specification of  \eqref{eq:R11_h},\eqref{eq:R1A_h}, \eqref{eq:trRAB_h} and \eqref{eq:ttRAB_h} for  a Bondi-Sachs metric \eqref{BSmetric}  the main equations are displayed in \cite[Eq. (28)-(30), (32)]{BSscolar},   (and correspond to those in \cite{newt} for $\lambda=1$).


\subsection{Summary of all Christoffel symboles}\label{sec:Christ}
\begin{subequations}
\begin{eqnarray}
\chr{0}{00} & = & 
 	(\ln  |g_{10}|)_{,0} - \f{V_{,1}}{2}  -\f{V}{2}(\ln |g_{01}|)_{,1}
\nonumber\\&&
	-\f{1}{2} g^{01}(g_{AB}U^AU^B)_{,1}\\
\chr{0}{01}&=& \chr{0}{11} = \chr{0}{1A} = 0\\
\chr{0}{0A} &=&
	\f{1}{2}g^{01}(D_A g_{01}+U_{A,1}) 	 \\
\chr{0}{BC} & = & 
	- \f{1}{2}g^{01}g_{BC,1} 
\end{eqnarray}	
\end{subequations}
\begin{subequations}
\begin{eqnarray}
\chr{1}{00} 
 & = & 
	 \f{1}{2}g^{01}(Vg_{01})_{,0}
	 + \f{1}{2}g^{01}(U_AU^A)_{,0}
	 - V(\ln  |g_{10}|)_{,0} 
\nonumber\\ 
&&
	 +\f{1}{2}g^{01}V(Vg_{01})_{,1}
	 +\f{1}{2}g^{01}V(U_AU^A)_{,1}
	 - g^{01}U^A U_{A,0}
\nonumber\\ 
&&	  
	  -\f{1}{2}g^{01}U^A D_A(Vg_{01})
	  -\f{1}{2}g^{01}U^AD_A(U_CU^C)\\
\chr{1}{01}
 & = &   \f{1}{2}g^{10}(Vg_{01})_{,1}
 	+\f{1}{2}g^{10}(U_AU^A)_{,1}
	-\f{1}{2}g^{10}U^AU_{A,1}
\nonumber\\ 
&&	  
	-\f{1}{2}U^AD_A \ln |g_{01}|
	\\	  
\chr{1}{0A} 
 & = & 
 	\f{1}{2} D_AV
 	+\f{1}{2}g^{10}D_A(U_CU^C)
	-\f{1}{2}g^{10}V U_{A,1}
\nonumber\\ 
&&	  
+\f{1}{2}g^{01}U^{B} g_{BA,0}
-g^{01}U^{B} U_{[A,B]}
\\
\chr{1}{11} & = &  (\ln|g_{01}|)_{,1} \\
 \chr{1}{1A}&=&
 	 \f{1}{2}g^{10}\Big(D_A g_{10}
	 -  g_{AB}U^B_{,1}\Big)\\
\chr{1}{AB} 
	& = &
	-g^{01}D_{(A}U_{B)}
	 -\f{1}{2}g^{10}\Big[
	g_{AB,0}
 	-Vg_{AB,1}\Big]	 
\end{eqnarray}	
\end{subequations}\begin{subequations}
\begin{eqnarray}
\chr{A}{00} & = & 
	\f{1}{2}g^{01}U^A  g_{10,0} 
     -\f{1}{2}g^{01}U^A(V g_{01}+ U_CU^C) _{,1}
\nonumber\\&&     
        - g^{AB} U_{B,0}
	 -\f{1}{2}D^B(V g_{01})
\nonumber\\&&     
	 -\f{1}{2}D^B(U_CU^C)	
\\
  \chr{C}{01}&=&
	- \f{1}{2}g^{CE} g_{EF,1}U^F- \f{1}{2}U^C_{,1}-\f{1}{2}D^C g_{01}  \\
\chr{A}{0B} & = &
	\f{1}{2}g^{01}\Big[ U^AD_B g_{01}
       +U^AU_{B,1}\Big]
\nonumber\\ 
&&	  
	 + \f{1}{2}g^{AC}g_{CB,0}
	 +g^{AC}D_{[C}U_{B]}	\\	
\chr{A}{11}&=&0\\
\chr{A}{1B} & = &  \f{1}{2}g^{AC}g_{CB,1}\\
\chr{A}{BC} 
 & = & 
 - \f{1}{2}g^{01}U^{A}g_{BC,1} 
 +\Upsilon^A_{BC}   
\end{eqnarray}
\end{subequations}

\begin{equation}
\label{ }
\Upsilon^A_{BC} :=\f{1}{2}g^{AE}(g_{BE,C}+g_{EC,B}-g_{BC,E})
\end{equation}


\section{Calculation of the Ricci tensor for the field equations for the general null metric $g_{ab}$}\label{sec_genRic}
For the calculation of the Ricci tensor we use the standard expression 
\begin{eqnarray}
\label{ }
R_{ab} &=& R^c_{\p{a}acb}\nonumber\\
&=& \chr{c}{ab,c} - (\ln \sqrt{-g})_{,ab} + \chr{c}{ab}(\ln \sqrt{-g})_{,c} 
\nonumber\\&&- \chr{c}{ad}\chr{d}{bc}\;,
\end{eqnarray}
for which we define the intermediate quantities
\begin{subequations}

\begin{eqnarray}
R^{(0)}_{ab} & = &\chr{c}{ab,c} \\
R^{(0)}_{ab} & = &- (\ln \sqrt{-g})_{,ab} \\
R^{(0)}_{ab} & = & \chr{c}{ab}(\ln \sqrt{-g})_{,c}\\
R^{(0)}_{ab} & = & - \chr{c}{ad}\chr{d}{bc}
\end{eqnarray}
\end{subequations}

\subsection{Calculation of $R_{00}$}

The intermediate variables  $R^{(a)}_{00}|_{x^1=0}$ are
\begin{subequations}
\begin{eqnarray}
R^{(1)}_{00}|_{x^1=0}
& = & 
(\ln  g_{10})_{,00} 
	 -  V_{,1}(\ln |g_{01}|)_{,0}
	 +\f{1}{2}(V_{,1})^2\\
R^{(2)}_{00}|_{x^1=0}&=&
 -(\ln\sqrt{-g})_{,00} \\
R^{(3)}_{00}|_{x^1=0}
	&=&
\Big[(\ln  g_{10})_{,0}  -\f{1}{2}V_{,1} \Big] (\ln\sqrt{-g})_{,0}\\
R^{(4)}_{00}|_{x^1=0}
	&=&
	- [(\ln  g_{10})_{,0} ]^2 
	+(\ln  g_{10})_{,0}V_{,1}
	 -\f{1}{2}[V_{,1}] ^2
\nonumber\\&&
		- \f{1}{4}g^{AC}g^{BD}g_{DA,0}g_{CB,0}
\end{eqnarray}
\end{subequations}
and their sum results to 
\begin{eqnarray}
R_{00}|_{x^1=0}
& = & 
	- V_{,1}\left (\ln\sqrt[4]{ |g_{01}|^2f} \right )_{,0}
	 -g_{01}\Big[\f{(\ln\sqrt{f})_{,0}}{g_{01}}\Big]_{,0}
\nonumber\\&&
	- \f{1}{4}g^{AC}g^{BD}g_{DA,0}g_{CB,0}\label{eq:R00_gen}
\end{eqnarray}
\begin{widetext}
\subsection{Calculation of $R_{0A}$}
The intermediate variables $R^{(a)}_{0A}$ are 
\begin{eqnarray}
\label{ }
R_{0A}^{(1)}|_{x^1=0}
 &=&
  	\f{1}{2}D_A (\ln | g_{01}|)_{,0}
	+\f{1}{2}[g^{01}g_{AB}U^B_{,1} ]_{,0}
	+ \f{1}{2}D_A V_{,1}
	-\f{1}{2} g^{10}g_{AB}V_{,1} U^B_{,1}
+\f{1}{2}g^{01}U^{B}_{,1} g_{BA,0}
	 +	  \f{1}{2}[g^{BC}g_{CA,0}]_{,B}
\\
\label{ }
R_{0A}^{(2)}|_{x^1=0}
 &=&
  -D_A(\ln|g_{01}|)_{,0}-(\ln\sqrt{f})_{,0A}\\
R_{0A}^{(3)}|_{x^1=0}
  &=& 
 	\f{1}{2}(D_A\ln | g_{01}|)(\ln |g_{01}|)_{,0}
	+\f{1}{2}(g^{01}U_{A,1})(\ln |g_{01}|)_{,0}
	+\f{1}{2}(D_A\ln | g_{01}|)(\ln \sqrt{f})_{,0}
\nonumber\\&&
		+\f{1}{2}(g^{01}U_{A,1})(\ln \sqrt{f})_{,0}
 +\f{1}{2}g^{CE}g_{EA,0} (\ln |g_{01}|)_{,C}
 +\f{1}{2}g^{CE}g_{EA,0} (\ln\sqrt{f})_{,C}\\
R_{0A}^{(4)}|_{x^1=0}
&=&
   - \f{1}{2}(\ln|g_{01}|)_{,0}(D_A\ln |g_{01}|)
   + \f{1}{2}V_{,1}  g^{01}U_{A,1}
   -\f{1}{2}g^{01}g_{AB}(\ln|g_{01}|)_{,0}(U^{B}_{,1})
	-\f{1}{2}g^{01}(U^C_{,1})(g_{AC,0})
\nonumber\\&&	
	-\f{1}{2}(D^C\ln g_{01})(g_{AC,0})
	- \f{1}{2}g^{CE}g_{DE,0}\Upsilon^{D}_{AC}
\end{eqnarray}
and adding them up yields
\begin{eqnarray}
R_{0A}|_{x^1=0} 
 &=&
\f{1}{2\sqrt{f}}[\sqrt{f}g^{01}U_{A,1}]_{,0}
 +	  \f{1}{2}D^C(g_{CA,0} )
 	- D_A\Big[(\ln\sqrt{f})_{,0}-\f{1}{2} V_{,1}\Big]
 	- \f{\sqrt{f}}{2} \Big(\f{D_A \ln | g_{01}|}{\sqrt{f}}\Big)_{,0}\label{eq:R0A_gen}
\end{eqnarray}
\end{widetext}
\subsection{Calculation of $R_{11}$}\label{sec:R_11}
The intermediate variables $R_{11}^{(a)}$
are
\begin{eqnarray}
R_{11}^{(1)} 
&=&  \chr{1}{11,1}
\\
R_{11}^{(2)}  &= &  -(\ln|g_{01}|)_{,11}-(\ln\sqrt{f})_{,11} 
\\
R_{11}^{(3)} 
&=&\chr{1}{11}(\ln|g_{01}|)_{,1}  
	+\chr{1}{11}(\ln\sqrt{f})_{,1} 
\\
R_{11}^{(4)} &=&
	-( \chr{1}{11})^2
	- \chr{C}{1D}\chr{D}{1C}
\end{eqnarray}
and its sum 
\begin{eqnarray}
R_{11} & = &  \chr{1}{11,1}
	-(\ln|g_{01}|)_{,11}
	-(\ln\sqrt{f})_{,11} 
\nonumber\\&&	
	+\chr{1}{11}(\ln|g_{01}|)_{,1}  
	+\chr{1}{11}(\ln\sqrt{f})_{,1}  
\nonumber \\&  & 
	-( \chr{1}{11})^2
	- \chr{C}{1D}\chr{D}{1C}
\end{eqnarray}
Insertion  of the Christoffel symboles $\chr{1}{11}$ and $\chr{A}{1B}$ gives after simplification
\begin{eqnarray}
R_{11} & = &  
	-(\ln\sqrt{f})_{,11} 
	+[(\ln|g_{01}|)_{,1}](\ln\sqrt{f})_{,1}  
\nonumber\\&&
	- \f{1}{4} g^{AC}g^{BD}g_{CB,1}g_{DA,1}\label{eq:R11_gen}
\end{eqnarray}
\subsection{Calculation of $R_{1A}$}
Calculation of the constituents gives
\begin{eqnarray}
R^{(1)}_{1A} & = & 
	\chr{1}{1A,1}+D_C \chr{C}{1A} - (\ln\sqrt{f})_{,E}\chr{E}{1A}
\nonumber\\&&	+ \Upsilon^E_{CA}\chr{C}{1E}
\\
R^{(2)}_{1A} & = & 
	D_A(\ln|g_{01}|)_{,1}
	+	(\ln\sqrt{f})_{,1A}
\\
R^{(3)}_{1A} & = & 
	\chr{1}{1A}(\ln\sqrt{-g})_{,1}
	+	\chr{C}{1A}(\ln|g_{01}|)_{,C}
\nonumber\\&&
	+	\chr{C}{1A}(\ln\sqrt{f})_{,C}
\\	
R_{1A}^{(4)} 
&=&
	- \chr{1}{11}\chr{1}{A1} 
	- \chr{1}{1C}\chr{C}{A1}
	- \chr{C}{10}\chr{0}{AC}
\nonumber\\&&
	- \chr{C}{1D}\chr{D}{AC}
\end{eqnarray}
so that
\begin{eqnarray}
R_{1A}  
& = & 
	\chr{1}{1A,1}
	+D_C \chr{C}{1A} 
	+ \Upsilon^E_{CA}\chr{C}{1E}
	-D_A(\ln|g_{01}|)_{,1}
\nonumber\\&&
	-	(\ln\sqrt{f})_{,1A}
	+\chr{1}{1A}(\ln|g_{01}|)_{,1}
	+\chr{1}{1A}(\ln\sqrt{f})_{,1}
\nonumber\\&&
	+	\chr{C}{1A}(\ln|g_{01}|)_{,C}
	- \chr{1}{11}\chr{1}{A1} 
	- \chr{1}{1C}\chr{C}{A1}
\nonumber\\&&
	- \chr{C}{10}\chr{0}{AC}
	- \chr{C}{1D}\chr{D}{AC}\label{eq:R_1A_1}
 \end{eqnarray}
Insert   $\chr{1}{11}$ 
and  $\chr{A}{BC}$ into \eqref{eq:R_1A_1}
  \begin{eqnarray}
R_{1A}  
& = & 
	\f{1}{\sqrt{f}}(\sqrt{f} \chr{1}{1A})_{,1}
	+D_C \chr{C}{1A}
	-D_A(\ln|g_{01}|)_{,1}
\nonumber\\&&
	-	(\ln\sqrt{f})_{,1A}
 	+	g^{01}\chr{C}{1A}D_Cg_{01} 
	- \chr{1}{1C}\chr{C}{A1}
\nonumber\\&&
	- \chr{C}{10}\chr{0}{AC}
	+  \f{1}{2}\chr{C}{1D}g^{01}U^{D}g_{AC,1} \;.\label{eq:R_1A_1}
 \end{eqnarray}
Now insert insert $\chr{0}{BC} $, $\chr{C}{01}$ and  $\chr{B}{1C}$ is the last term 
  \begin{eqnarray}
R_{1A}  
& = & 
	\f{1}{\sqrt{f}}(\sqrt{f} \chr{1}{1A})_{,1}
	+D_C \chr{C}{1A}
	-D_A(\ln|g_{01}|)_{,1}
\nonumber\\&&
	-	(\ln\sqrt{f})_{,1A}
 	+\f{1}{2}	g^{01}g^{CE}g_{AE,1} D_Cg_{01} 
\nonumber\\&&
	-\f{1}{2} \chr{1}{1C}g^{CE}g_{AE,1}
 - \f{1}{4}U^C_{,1}g^{01}g_{AC,1}
\nonumber\\&&
 -\f{1}{4}(D^C g_{01})g^{01}g_{AC,1}  	\label{eq:R_1A_2}
\end{eqnarray}
Using $\chr{1}{1A}$ in \eqref{eq:R_1A_2} gives
  \begin{eqnarray}
R_{1A}  
& = & 
	-\f{\sqrt{f} }{2} \Big[\f{ D_A\ln|g_{10}|}{\sqrt{f}}\Big]_{,1}
	-\f{1}{2\sqrt{f}}(\sqrt{f} g^{10} g_{AB}U^B_{,1})_{,1}
\nonumber\\&&
	+D_C \chr{C}{1A}
	-	(\ln\sqrt{f})_{,1A}	
  \end{eqnarray}
after  simplification and inserting   $\chr{A}{1B}$ yields the final expression
  \begin{eqnarray}
R_{1A}  &=&
	-\f{1}{2\sqrt{f}}(\sqrt{f} g^{10} g_{AB}U^B_{,1})_{,1}
	-\f{\sqrt{f} }{2} \Big[\f{ D_A\ln|g_{10}|}{\sqrt{f}}\Big]_{,1}
\nonumber\\&&
	-	(\ln\sqrt{f})_{,1A}	
	+\f{1}{2}D^C g_{CA,1}\;\;.\label{eq:R1A_gen}
  \end{eqnarray}

\subsection{Calculation of $R_{AB}$}  
The intermediate variables for the Ricci tensor are
\begin{eqnarray}
R^{(1)}_{AB} &=&   \chr{0}{AB,0}+\chr{1}{AB,1}+\chr{C}{AB,C}
\\
R^{(2)}_{AB} &=&  -(\ln|g_{01}|)_{,AB}-(\ln\sqrt{f})_{,AB}
\\
R^{(3)}_{AB} &=& 
	\chr{0}{AB}(\ln\sqrt{-g})_{,0} + \chr{1}{AB}(\ln\sqrt{-g})_{,1} 
	\nonumber\\&&
	+ \chr{C}{AB}(\ln|g_{01}|)_{,C} 
	+ \chr{C}{AB}(\ln\sqrt{f})_{,C} 
\\
R^{(4)}_{AB} &=& 
	  -  \chr{0}{A0}\chr{0}{B0} 
         -  \chr{0}{AC}\chr{C}{B0} 
	- \chr{1}{A1}\chr{1}{B1} 
	- \chr{1}{AC}\chr{C}{B1} 
\nonumber\\&&
	- \chr{C}{A0}\chr{0}{BC}
	- \chr{C}{A1}\chr{1}{BC}
	- \chr{C}{AD}\chr{D}{BC}
\end{eqnarray}
which give
\begin{eqnarray}
R_{AB} 
 & = & 
	\chr{0}{AB,0}+\chr{1}{AB,1}+\chr{C}{AB,C}
	 -(\ln|g_{01}|)_{,AB}
\nonumber
\\&&
	 -(\ln\sqrt{f})_{,AB}
	+\chr{0}{AB}(\ln\sqrt{-g})_{,0} 
	+ \chr{1}{AB}(\ln\sqrt{-g})_{,1} 
\nonumber\\&&
	+ \chr{C}{AB}(\ln|g_{01}|)_{,C} 
 + \chr{C}{AB}(\ln\sqrt{f})_{,C} 
	-  \chr{0}{A0}\chr{0}{B0} 
\nonumber\\&&
         -  \chr{0}{AC}\chr{C}{B0} 
	- \chr{1}{A1}\chr{1}{B1} 
	- \chr{1}{AC}\chr{C}{B1} 
\nonumber\\&&
	- \chr{C}{A0}\chr{0}{BC}
	- \chr{C}{A1}\chr{1}{BC}
	- \chr{C}{AD}\chr{D}{BC}
\end{eqnarray}
We combine $\chr{0}{AB,0}$ and $\chr{0}{AB,1}$ with derivatives of $\sqrt{-g}$
\begin{eqnarray}
R_{AB} 
 & = & 
\f{1}{\sqrt{-g}}\Big(\sqrt{-g} \chr{0}{AB,0}\Big)_{,0}
	+\f{1}{\sqrt{-g}}\Big(\sqrt{-g} \chr{0}{AB,0}\Big)_{,1}
\nonumber\\&&
	+\chr{C}{AB,C}
	 -(\ln|g_{01}|)_{,AB}-(\ln\sqrt{f})_{,AB}
\nonumber\\&&
	+ \chr{C}{AB}D_C(\ln|g_{01}|)
 + \chr{C}{AB}D_C(\ln\sqrt{f})
\nonumber\\&&
	-  \chr{0}{A0}\chr{0}{B0} 
         -  \chr{0}{AC}\chr{C}{B0} 
	- \chr{1}{A1}\chr{1}{B1} 
	- \chr{1}{AC}\chr{C}{B1} 
\nonumber\\&&
	- \chr{C}{A0}\chr{0}{BC}
	- \chr{C}{A1}\chr{1}{BC}
	- \chr{C}{AD}\chr{D}{BC}\;\;.
\end{eqnarray}
Insertion of  $\chr{A}{BC}$, 
with using the  cov derivatives 
\begin{eqnarray}
D_AD_B \ln|g_{01}| &=& (D_B \ln|g_{01}|)_{,A}-\Upsilon^{C}_{AB}D_C( \ln|g_{01}|)\nonumber\\
&&\\
D_C\Big(U^{C}g_{AB,1}\Big)
	&=&
	\Big(U^{C}g_{AB,1}\Big)_{,C}
	+\big(U^{C}g_{AB,1}\big)	(\ln\sqrt{f})_{,C} 
\nonumber          \\
&&          - \Upsilon^E_{AC}U^{C}g_{BE,1}
          -\Upsilon^E_{BC}U^{C}g_{AE,1}, 
\end{eqnarray}
the relation
$
D_Cg^{01} 
= -g^{01}D_C\ln|g_{01}|
$
and the definitions
\begin{eqnarray}
R^{(2)}_{AB} & = &               
           \Upsilon^C_{AB,C}
	 -(\ln\sqrt{f})_{,AB}
	+\Upsilon^C_{AB}(\ln\sqrt{f})_{,C} 
\nonumber\\&&
     - \Upsilon^E_{AC}\Upsilon^C_{BE}
     \\
\Gamma_{AB}&=&     
	-  \chr{0}{A0}\chr{0}{B0} 
	- \chr{1}{A1}\chr{1}{B1} 
       - 2 \chr{0}{C(A}\chr{C}{B)0} 
\nonumber\\&&
	- 2\chr{1}{C(A}\chr{C}{B)1} 
\end{eqnarray}
gives
\begin{eqnarray}
R_{AB} & = & 
	 R^{(2)}_{AB}
	 -D_AD_B \ln|g_{01}|
	+ \f{1}{\sqrt{-g}}\Big(\sqrt{-g} \chr{0}{AB,0}\Big)_{,0}
\nonumber\\&&
	+\f{1}{\sqrt{-g}}\Big(\sqrt{-g} \chr{0}{AB,0}\Big)_{,1}
	- \f{g^{01}}{2}\Big(D_C U^{C}g_{AB,1}\Big)
\nonumber\\&&
	-  \f{1}{4}(g^{01})^2U^{C}U^{E}g_{AE,1}g_{BC,1}
	+\Gamma_{AB}	\;\;,
\end{eqnarray}
where $R^{(2)}_{AB}$ is the Ricci tensor w.r.t. the 2-metric $g_{AB}$.
To determine $\Gamma_{AB}$, we calculate  
\begin{widetext}
\begin{eqnarray}
\lefteqn{	-  \chr{0}{A0}\chr{0}{B0} 
	- \chr{1}{A1}\chr{1}{B1} 
	=
-\f{1}{2}(g^{01})^2\Big[(D_A g_{01})(D_B g_{01} )
	+  U^Eg_{E(A,1}D_{B)} g_{01} 
	+g_{F(A}g_{B)E,1}U^E_{,1}U^F}
&&\nonumber\\&&
	+\f{1}{2}U^EU^Fg_{EA,1}g_{BF,1}	
	+U^E_{,1}U^F_{,1}g_{EA}g_{BF}	
	\Big]\\
\lefteqn{ -2 \chr{C}{0(B}\chr{0}{A)C} -2 \chr{C}{1(B}\chr{1}{A)C} 
=
	\f{1}{2}(g^{01})^2\Big[ U^Cg_{C(A,1} D_{B)} g_{01}
	+U^C U^Eg_{C(A,1} g_{B)E,1}
	+U^CU^E_{,1} g_{C(A,1} g_{B)E}\Big]}\nonumber&&\\
\nonumber\\&&	 
	 +\f{1}{2}g^{01}g^{CE}\Big[ g_{CA,1} D_{E}U_{B}
		 + g_{CB,1} D_{E}U_{A}\Big]
	 + g^{01}g^{CE} g_{C(A,1} g_{B)E,0}
 	-\f{1}{2}Vg^{10}g^{CE}g_{E(A,1}g_{B)C,1}
\end{eqnarray}
Inserting the resulting $\Gamma_{AB}$ and $\sqrt{-g} $ into $R_{AB}$ gives
\begin{eqnarray}
R_{AB}
& = & 
	R^{(2)}_{AB}
-\f{2}{\sqrt{|g_{01}|}}D_AD_B \sqrt{|g_{01}|}
	- \f{1}{2}g^{01}D_C\Big(U^{C}g_{AB,1}\Big)
	+\f{1}{|g_{01}|\sqrt{f}}\Big(|g_{01}|\sqrt{f}\chr{0}{AB}\Big)_{,0}
	+\f{1}{|g_{01}|\sqrt{f}}\Big(|g_{01}|\sqrt{f}\chr{1}{AB}\Big)_{,1}
\nonumber\\&&
	-\f{1}{2}(g^{01})^2U^E_{,1}U^F_{,1}g_{EA}g_{BF}	
	 + g^{01}g^{CE} g_{C(A,1} g_{B)E,0}
	 +\f{1}{2}g^{01}g^{CE}\Big[ g_{CA,1} D_{E}U_{B}
		 + g_{CB,1} D_{E}U_{A}\Big]
\nonumber
\\&&
 	-\f{1}{2}Vg^{10}g^{CE}g_{E(A,1}g_{B)C,1}\label{eq:R_AB_1}
\end{eqnarray}
where we used
\begin{equation}
\label{ }
-\f{2}{\sqrt{|g_{01}|}}D_AD_B \sqrt{|g_{01}|}
=
 -D_AD_B \ln|g_{01}|
	 -\f{1}{2} (D_A \ln|g_{01}|)(D_B \ln|g_{01}| )\;.
\end{equation}
Inserting  $\chr{0}{AB}$,
$\chr{1}{AB}$ and simplification of \eqref{eq:R_AB_1} gives
\begin{eqnarray}
R_{AB}
& = & 
R^{(2)}_{AB}
-\f{2}{\sqrt{|g_{01}|}}D_AD_B \sqrt{|g_{01}|}
	- \f{1}{2}g^{01}D_C\Big(U^{C}g_{AB,1}\Big)
	-\f{g^{01}}{2\sqrt{f}}\bigg\{\Big[\sqrt{f} g_{AB,1}\Big]_{,0}
	+\Big[\sqrt{f}g_{AB,0}\Big]_{,1}\bigg\}
\nonumber\\&&
	-\f{g^{01}}{\sqrt{f}}\Big[\sqrt{f}
	   D_{(A}U_{B)} \Big]_{,1}
 	+\f{g^{01}}{2\sqrt{f}}\Big[\sqrt{f}Vg_{AB,1}\Big]_{,1}
	 	-\f{1}{2}Vg^{10}g^{CE}g_{E(A,1}g_{B)C,1}
-\f{1}{2}(g^{01})^2\Big[ 
	U^E_{,1}U^F_{,1}g_{EA}g_{BF}	
	\Big]
\nonumber
\\&&
	 + g^{01}g^{CE} g_{C(A,1} g_{B)E,0}
	 +\f{1}{2}g^{01}g^{CE}\Big[ g_{CA,1} D_{E}U_{B}
	 + g_{CB,1} D_{E}U_{A}\Big]\;\;.
\end{eqnarray}
Since
\begin{eqnarray}
(\sqrt{f} g_{AB,1})_{,0}
	+(\sqrt{f}g_{AB,0})_{,1}
 & = & 
 	2 (\sqrt{f}) g_{AB,10}
	 + (\sqrt{f})_{,0} g_{AB,1}
	 + (\sqrt{f})_{,1} g_{AB,0}
\end{eqnarray}
we have 
\begin{eqnarray}
R_{AB}
& = & 
R^{(2)}_{AB}
	-g^{01}\bigg\{ g_{AB,10}
	 +\f{1}{2}\Big[ (\ln\sqrt{f})_{,0} g_{AB,1}
	 + (\ln \sqrt{f})_{,1} g_{AB,0}\Big]
	 -g^{CE} g_{C(A,1} g_{B)E,0}\bigg\}
-\f{2}{\sqrt{|g_{01}|}}D_AD_B \sqrt{|g_{01}|}
\nonumber
\\&&
 	+\f{g^{01}}{2}\bigg\{\Big[Vg_{AB,1}\Big]_{,1}
 	+V\Big[(\ln\sqrt{f})_{,1}g_{AB,1}
	 	-g^{CE}g_{E(A,1}g_{B)C,1}\Big]\bigg\}
-\f{1}{2}(g^{01})^2\Big[ 
	U^E_{,1}U^F_{,1}g_{EA}g_{BF}	
	\Big]
\nonumber\\&&
	- \f{1}{2}g^{01}\bigg\{U^{C}D_Cg_{AB,1}
	+(D_C U^{C})g_{AB,1}
	 -g^{CE}\Big(D_{E}U_{A} \Big)g_{BC,1} 
	 -g^{CE}\Big(D_{E}U_{B} \Big)g_{AC,1} 	 	
	+\f{2}{\sqrt{f}}\Big[\sqrt{f}
	   D_{(A}U_{B)} \Big]_{,1}
	   \bigg\}\;\;.
\nonumber\\&&
\end{eqnarray}
Rearranging and 
multiplication with $-g_{01}$, while using $g_{01} = \epsilon|g_{01}|$,  gives
\begin{eqnarray}
0&=&
	-g_{01}R^{(2)}_{AB}
+2\epsilon\sqrt{|g_{01}|}D_AD_B \sqrt{|g_{01}|}
	+ g_{AB,10}
	 +\f{1}{2}\Big[ (\ln\sqrt{f})_{,0} g_{AB,1}
	 + (\ln \sqrt{f})_{,1} g_{AB,0}\Big]
	 -g^{CE} g_{C(A,1} g_{B)E,0}
\nonumber
\\&&
 	-\f{1}{2}\bigg\{\Big[Vg_{AB,1}\Big]_{,1}
 	+V\Big[(\ln\sqrt{f})_{,1}g_{AB,1}
	 	-g^{CE}g_{E(A,1}g_{B)C,1}\Big]\bigg\}
	+ \f{1}{2}U^{C}D_Cg_{AB,1}
	+\f{1}{2}(D_C U^{C})g_{AB,1}
\nonumber\\&&
	 -\f{1}{2}g^{CE}\Big[\Big(D_{E}U_{A} \Big)g_{BC,1} 
	+\Big(D_{E}U_{B} \Big)g_{AC,1} \Big]	 	
	+(\ln\sqrt{f})_{,1} D_{(A}U_{B)} 
	+[ D_{(A}U_{B)}]_{,1}
+\f{1}{2}(g^{01})\Big[ 
	U^E_{,1}U^F_{,1}g_{EA}g_{BF}	
	\Big]
\nonumber\\&&
 + g_{01}R_{AB}\;\;.\label{eq:RAB_intermediate}
\end{eqnarray}
Since
\begin{eqnarray}
(D_AU_C)_{,1} 
 & = &   D_A U_{C,1}-U^F\Big( \Upsilon_{CAF,1}- g_{FH,1}\Upsilon^H_{CA} \Big)
\end{eqnarray}
and 
\begin{eqnarray}
\Upsilon_{CAF,1} 
&=&
	\f{1}{2}(D_C g_{FA,1} +D_A g_{CF,1} -D_F g_{CA,1})
	+ \Upsilon^{E}_{AC}g_{FE,1}
\end{eqnarray}
we have 
\begin{equation}
\label{ }
\big[D_{(A}U_{B)}\big]_{,1} 
=D_{(A}  U_{B),1}
 -U^F D_{(A}g_{B)F,1} +\f{1}{2}U^FD_F g_{AB,1}\;,
\end{equation}
so that insertion into \eqref{eq:RAB_intermediate}
gives
\begin{eqnarray}
0&=&
-g_{01}R^{(2)}_{AB}
+2\epsilon\sqrt{|g_{01}|}D_AD_B \sqrt{|g_{01}|}
	+ g_{AB,10}
	 +\f{1}{2}\Big[ (\ln\sqrt{f})_{,0} g_{AB,1}
	 + (\ln \sqrt{f})_{,1} g_{AB,0}\Big]
	 -g^{CE} g_{C(A,1} g_{B)E,0}
\nonumber
\\&&
 	-\f{1}{2}\bigg\{\Big[Vg_{AB,1}\Big]_{,1}
 	+V\Big[(\ln\sqrt{f})_{,1}g_{AB,1}
	 	-g^{CE}g_{E(A,1}g_{B)C,1}\Big]\bigg\}
	+ U^{C}D_Cg_{AB,1}
	+\f{1}{2}(D_C U^{C})g_{AB,1}
\nonumber\\&&
	 - g_{F(A}g_{B)C,1} \Big(D^CU^F - D^FU^C \Big)
	+(\ln\sqrt{f})_{,1} D_{(A}U_{B)} 
	+ g_{E(A} D_{B)}  U^E_{,1}
	+\f{1}{2}(g^{01})\Big[ 
	U^E_{,1}U^F_{,1}g_{EA}g_{BF}	
	\Big] 
\nonumber\\&&
	+ g_{01}R_{AB} \label{eq:R_AB_final_ok}
\end{eqnarray}
where we combined and simplified some terms, revoked the some of the notation $U_{A} = g_{AB}U^B$  and also introduced the shorthand notation $D^A = g^{AB}D_B$. 

\subsubsection{Calculation of $g_{01}g^{AB}R_{AB}$}
We contract  \eqref{eq:R_AB_final_ok} with $g^{AB}$ to find
\begin{eqnarray}
\lefteqn{0
 = 
-g_{01}g^{AB}R^{(2)}_{AB}
+2\epsilon\sqrt{|g_{01}|}D^AD_A \sqrt{|g_{01}|}
	+g^{AB} g_{AB,10}
	 +\f{1}{2}g^{AB}\Big[ (\ln\sqrt{f})_{,0} g_{AB,1}
	 + (\ln \sqrt{f})_{,1} g_{AB,0}\Big]
	}&&
\nonumber\\&&
	 -g^{CE} g_{C(A,1} g_{B)E,0}
 	-\f{1}{2}g^{AB}\bigg\{\Big[Vg_{AB,1}\Big]_{,1}
 	+V\Big[(\ln\sqrt{f})_{,1}g_{AB,1}
	 	-g^{CE}g_{E(A,1}g_{B)C,1}\Big]\bigg\}
\nonumber\\&&
	+2 U^{C}D_C(\ln\sqrt{f})_{,1}
	+2(D_C U^{C})(\ln\sqrt{f})_{,1}
	+  D_{E}  U^E_{,1}
	+\f{1}{2}(g^{01})\Big[ 	U^E_{,1}U^F_{,1}g_{EF}	
	\Big] 
	+ g_{01}g^{AB}R_{AB} \label{eq:gAB_RAB_1}
\end{eqnarray}
As we have for $ \alpha=0,1$
\begin{equation}
\label{ }
g^{AB}g_{AB,\alpha}  
= 2  (\ln \sqrt{f})_{,\alpha}
\end{equation}
as well as 
\begin{eqnarray}
\label{ }
2 U^{C}D_C(\ln\sqrt{f})_{,1}
	+2(D_C U^{C})(\ln\sqrt{f})_{,1}
&=&
	D_C[f^{-1} ( f U^{C})_{,1}]
	-D_C U^{C}_{,1} 
\end{eqnarray}
\eqref{eq:gAB_RAB_1} can be written as
\begin{eqnarray}
0&=&-g_{01}g^{AB}R^{(2)}_{AB}
+2\epsilon\sqrt{|g_{01}|}D^AD_A \sqrt{|g_{01}|}
	+g^{AB} g_{AB,10}
	 +2 (\ln\sqrt{f})_{,0}(\ln\sqrt{f})_{,1} 
	 -g^{CE} g_{C(A,1} g_{B)E,0}
\nonumber
\\&&
 	-\f{1}{2}g^{AB}\bigg\{\Big[Vg_{AB,1}\Big]_{,1}
 	+V\Big[(\ln\sqrt{f})_{,1}g_{AB,1}
	 	-g^{CE}g_{E(A,1}g_{B)C,1}\Big]\bigg\}
	+D_C\Big[f^{-1} ( f U^{C})_{,1}\Big]
	+\f{1}{2}(g^{01})\Big[ 	U^E_{,1}U^F_{,1}g_{EF}	
	\Big] 
\nonumber\\&&
	+ g_{01}g^{AB}R_{AB}\label{eq:gAB_R_AB_2}
\end{eqnarray}
Inserting
\begin{eqnarray}
 g^{AB}g_{AB,01} = 2  (\ln \sqrt{f})_{,01}  + g^{AC}g^{BD}g_{CD,0}g_{AB,1}
\end{eqnarray}
into \eqref{eq:gAB_R_AB_2} results into
\begin{eqnarray}
\lefteqn{0
 = 
-g_{01}g^{AB}R^{(2)}_{AB}
+2\epsilon\sqrt{|g_{01}|}D^AD_A \sqrt{|g_{01}|}
	+2  (\ln \sqrt{f})_{,01} 
	 + g^{AC}g^{BD}g_{CD,0}g_{AB,1}
	  +2 (\ln\sqrt{f})_{,0}(\ln\sqrt{f})_{,1} 
}&&
\nonumber\\&&
	 -g^{CE} g_{CA,1} g_{BE,0}
 	-\f{1}{2}g^{AB}\bigg\{\Big[Vg_{AB,1}\Big]_{,1}
 	+V\Big[(\ln\sqrt{f})_{,1}g_{AB,1}
	 	-g^{CE}g_{EA,1}g_{BC,1}\Big]\bigg\}
	+D_C\Big[f^{-1} ( f U^{C})_{,1}\Big]
\nonumber\\&&	
	+\f{1}{2}(g^{01})\Big[ 	U^E_{,1}U^F_{,1}g_{EF}	
	\Big] 
	+ g_{01}g^{AB}R_{AB}
\end{eqnarray}
Simplification and modification of terms containing $V$, 
while using  $g^{AB}_{,1} = -g^{AC}g^{BD}g_{CD,1}$ and the  derivative of  {a determinant} yields 
\begin{eqnarray}
0&=&
-g_{01}g^{AB}R^{(2)}_{AB}
+2\epsilon\sqrt{|g_{01}|}D^AD_A \sqrt{|g_{01}|}
	+\f{2}{\sqrt{f}} \Big[ \sqrt{f} (\ln \sqrt{f})_{,0}\Big]_{,1} 
 	-\Big[V(\ln \sqrt{f})_{,1}\Big]_{,1}
	+D_C\Big[f^{-1} ( f U^{C})_{,1}\Big]
\nonumber
\\&&
 	-\f{1}{2}\bigg\{-Vg^{AB}_{,1} g_{AB,1} 
 	+V\Big[2(\ln\sqrt{f})_{,1}(\ln \sqrt{f})_{,1}
	 	+(g^{CB})_{,1}g_{BC,1}\Big]\bigg\}
	+\f{1}{2}(g^{01})\Big[ 	U^E_{,1}U^F_{,1}g_{EF}	
	\Big] 
	+ g_{01}g^{AB}R_{AB}\;\;,
	\nonumber\\
\end{eqnarray}
whose further simplification gives
after  multiplication with $\sqrt{f}$
\begin{eqnarray}
0&=&
-g_{01}\sqrt{f} g^{AB}R^{(2)}_{AB}
+2\epsilon\sqrt{|g_{01}|}\sqrt{f} D^AD_A \sqrt{|g_{01}|}
	+ \Big[ \f{f_{,0}}{\sqrt{f}}\Big]_{,1} 
 	-\f{1}{2}\Big[\f{Vf_{,1}}{ \sqrt{f}}\Big]_{,1}
	+\sqrt{f}D_C\Big[f^{-1} ( f U^{C})_{,1}\Big]
\nonumber\\&&
	+\f{1}{2}(g^{01})\sqrt{f}g_{EF}	U^E_{,1}U^F_{,1}	
	+ g_{01}\sqrt{f} g^{AB}R_{AB}\;\;.\label{eq:trRAB_gen}
\end{eqnarray}
\subsubsection{Calculation of $g_{01}t^A t^BR_{AB}$}
The properties of the 2-dyad $t^A$ defined in \eqref{eq:def_tA} imply
\begin{subequations}\label{eq:prop_tA}
\begin{eqnarray}
 t^{A}\bar t^{B}g_{AB,a} &=& (\ln f)_{,a}\\
t^At^Bg^{CE} g_{CA,1} g_{BE,0} 
	& = &
	t^A t^{B}\Big[g_{AB,1} (\ln\sqrt{ f} )_{,0}
         +g_{AB,0} (\ln\sqrt f )_{,1}\Big]         \\
t^At^Bg^{CE} g_{C(A,1} g_{B)E,1} 
	 &=&  2t^A t^{B} g_{AB,1} (\ln\sqrt f )_{,1}         
\end{eqnarray}
\end{subequations}
We contract \eqref{eq:R_AB_final_ok} with $t^At^B$, insert \eqref{eq:prop_tA} and  simplify to find
\begin{eqnarray}
0&=&
  t^At^B\Big\{
-g_{01}R^{(2)}_{AB}
+2\epsilon \sqrt{|g_{01}|}D_AD_B \sqrt{|g_{01}|}
	+ g_{AB,10} -\f{1}{2}\Big[ (\ln\sqrt{f})_{,0} g_{AB,1}
	 + (\ln \sqrt{f})_{,1} g_{AB,0}\Big]\Big\}
\nonumber
\\&&
 	+t^At^B\bigg\{-\f{\sqrt{f}}{2}\Big[\f{Vg_{AB,1}}{\sqrt{f}}\Big]_{,1}  
	+U^{C}D_Cg_{AB,1}
	+\f{1}{2}(D_C U^{C})g_{AB,1}
	 - g_{FA}g_{BC,1} \Big(D^CU^F - D^FU^C \Big)
	 \Big\}
\nonumber\\&&
	+t^At^B\Big\{(\ln\sqrt{f})_{,1} D_{A}U_{B} 
	+g_{EA} D_{B}  U^E_{,1}
	+\f{1}{2}(g^{01})\Big[ U^E_{,1}U^F_{,1}g_{EA}g_{BF}	
	\Big] 
		+ g_{01}R_{AB} 
	\Big\}\;\;.
	\nonumber\\
\end{eqnarray}
From the properties of the Riemann tensor, we deduce
\begin{eqnarray}
\label{ }
t^At^BR^{(2)}_{AB}& =&t^At^BR^{(2)C}_{\p{(2)C}ACB} =t^At^Bt^{(C}\bar t^{D)} R^{(2)}_{DACB}
=
\f{1}{2}(t^At^Bt^{C}\bar t^{D} R^{(2)}_{DACB}+t^At^Bt^{D}\bar t^{C} R^{(2)}_{DACB})
=0\;\;,
\end{eqnarray}
so that
\begin{eqnarray}
0&=&
 t^At^B\Bigg\{
2\epsilon \sqrt{|g_{01}|}D_AD_B \sqrt{|g_{01}|}
 	 -\f{\sqrt{f}}{2}\Big[\f{Vg_{AB,1}}{\sqrt{f}}\Big]_{,1} 
	 + g_{AB,10} -\f{1}{2}\Big[ (\ln\sqrt{f})_{,0} g_{AB,1}
	 + (\ln \sqrt{f})_{,1} g_{AB,0}\Big]
\nonumber\\&&
	\qquad \;\;+U^{C}D_Cg_{AB,1}
	+\f{1}{2}(D_C U^{C})g_{AB,1}
	+(\ln\sqrt{f})_{,1} D_{A}U_{B} 
	+g_{EA} D_{B}  U^E_{,1}
		 - g_{FA}g_{BC,1} \Big(D^CU^F - D^FU^C \Big)
\nonumber\\&&
	\qquad\;\;+\f{1}{2}(g^{01})\Big[ U^E_{,1}U^F_{,1}g_{EA}g_{BF}
			\Big] 
		+ g_{01}R_{AB} \label{eq:ttRAB_gen}
	\Bigg\}\;\;.
\end{eqnarray}
\end{widetext}


\section{Calculation of the Ricci tensor for the field equations with the conformal decomposition of $g_{AB}$}\label{sec_confRic}
For  the conformal decomposition $g_{AB} = r^2h_{AB}$, we list some  relevant relations used to simplify the resulting expressions.  
From  $g_{AB} = r^2h_{AB}$ with $g^{AB} =r^{-2}h^{AB} $, we deduce the volume element
\begin{equation}
\label{ }
\sqrt f =r^2\sqrt{h }\;\;,
\end{equation}
 where $h(x^A) = \det(h_{AB}) $.
The Christoffel symbols transform under the conformal transformation
\begin{equation}
\label{ }
\Upsilon^C_{AB}(r^2h_{AB}) = H^C_{AB}
+  \f{1}{r} \Big(\kron{C}{A}\mc{D}_{B} r+\kron{C}{B}\mc{D}_{A} r -  h_{AB}\mc{D}^C  r\Big)
\end{equation}
where
\begin{equation}
\label{ }
H^C_{AB}:= \f{1}{2}h^{CD}(h_{AD,B}+h_{AD,B} - h_{AB,D})\;\;.
\end{equation}
is the Christoffel symbol with respect to $h_{AB}$. 
We denote the covariant derivative wrt $h_{AB}$ with $\mc{D}_A$ and set $\mc{D}^A = h^{AB}\mc{D}_B$. This gives $D^A = r^{-2}\mc{D}^A$.
The Ricci tensor ${R}^{(2)}_{AB}$ transforms as 
\begin{equation}
\label{ }
{R}^{(2)}_{AB} = \mc{R}_{AB} - h_{AB} (h^{EF}\mc{D}_E\mc{D}_F \ln r) 
\end{equation}
where $\mc{R}_{AB}$ is the Ricci tensor wrt $h_{AB}$ whose contraction with $h^{AB}$ gives
\begin{equation}
\label{eq:conf_RS}
h^{AB}{R}^{(2)}_{AB} = \mc{R}- \mc{D}^F\mc{D}_F \ln r^2\;\;.
\end{equation}
where $ \mc{R}$ is the Ricci scalar wrt $h_{AB}$.
Also,  have for 2-vector  {fields, $U^A$,}
\begin{eqnarray}
D_E U^A 
& = & 
	\mc{D}_E U^A + U^F\kron{A}{(E}\mc{D}_{F)}\ln r^2 -  h_{FE}U^F\mc{D}^A \ln r \nonumber\\	
&&\\	
D_E U^E 
 & = &
\f{1}{r^2}\mc{D}_E \Big( r^2 U^E\Big)
\end{eqnarray}
for co-vector  {fields,  $X_A$, }
\begin{eqnarray}
D_E X_F & = & 
	\mc{D}_E X_F -X_H\kron{H}{(E}\mc{D}_{F)}\ln r^2
	\nonumber\\&&
	+ h_{FE}(\mc{D}^H \ln r)X_H 
\end{eqnarray}
and covariant symmetric  {2-tensors, $X_{AB}$,}
\begin{eqnarray}
D_C X_{AB} 
& = & 
	r^2\mc{D}_C\Big(\f{ X_{AB} }{r^2}\Big)
	-X_{C(A}\mc{D}_{B)}\ln r^2
\nonumber\\&&
	 { +  h_{C(A}X_{B)H}\mc{D}^H \ln r^2}\\
h^{CB}D_C X_{AB}  & = & 
	 \mc{D}^B X_{AB}  
	 	-  (h^{CB}X_{CB})\mc{D}_{A}\ln r 
\end{eqnarray}
For the coordinates $x^e = (x^0, x^1)$,  consider
\begin{eqnarray}
\lefteqn{r^4 g^{AC}g^{BD}g_{CB,e}g_{DA,f}
=  
 2r^4  (\ln r^2)_{,e}(\ln r^2)_{,f}  }&&
 \nonumber\\&&
 + r^4 h_{CB,e}h_{DA,f}h^{AC}h^{BD}\qquad\qquad
 \label{trgAB,a_quad}
\end{eqnarray}
and 
 {
\begin{eqnarray}
	D^B g_{AB,e}	& = &
	D_C[(\ln r^2)_{,e}\kron{C}{A}] + g^{BC}D_B (r^2h_{AC,e})
	\\
	&=&
	\mc{D}_A (\ln r^2)_{,e} 
	+(\mc{D}^E \ln r^2)h_{AE,e}
\nonumber\\&&	+\mc{D}^B h_{AB,e} \;\;.
	 \label{DBg_AB,a}
\end{eqnarray}
}
\subsection{$R_{00}$}
Using \eqref{eq:conf_gAB} and \eqref{trgAB,a_quad} in \eqref{eq:R00_gen} gives
\begin{eqnarray}
\lefteqn{R_{00}|_{x^1=0}
 = 
	- V_{,1} (\ln r  )_{,0}
	 -g_{01}\Big[\f{(\ln r^2)_{,0}}{g_{01}}\Big]_{,0}}&&
\nonumber\\&&
	  -\f{1}{2}[ (\ln r^2)_{,0}]^2
	  -\f{1}{4} h_{CB,0}h_{DA,0}h^{AC}h^{BD},
	  \label{eq:R00_h}
\nonumber\\	  
\end{eqnarray}
which becomes after specification to the affine, null metric \eqref{eq:spec_anm}
\begin{eqnarray}
R_{00}|_{x^1=0}
& = & 
	 - \f{2r_{,00}}{r}  
	  -\f{1}{4} h_{CB,0}h_{DA,0}h^{AC}h^{BD}\;\;.
\nonumber\\	  
\end{eqnarray}
where the right hand side is evaluated at $x^1=0$.
\subsection{$R_{0A}$}
Using \eqref{eq:conf_gAB} and \eqref{DBg_AB,a} in \eqref{eq:R0A_gen} gives
\begin{eqnarray}
\lefteqn{R_{0A}|_{x^1=0} 
=
\f{[ r^4 g^{01}h_{AB}U^B_{,1}]_{,0}}{2r^2}
 	- \f{r^2}{2} \Big(\f{\mc{D}_A \ln | g_{01}|}{r^2}\Big)_{,0}
}\nonumber&&
 \nonumber\\
 && 
	 {+\f{1}{2}\Big[\mc{D}_A (\ln r^2)_{,0} }
	+(\mc{D}^E \ln r^2)h_{AE,0}
\nonumber\\&&	+\mc{D}^B h_{AB,0} \Big]
 	- \mc{D}_A\Big[(\ln r^2)_{,0}-\f{1}{2} V_{,1}\Big]
 \label{eq:R0A_h}
\end{eqnarray}
again specification to \eqref{eq:spec_anm}
\begin{eqnarray}
\lefteqn{R_{0A}|_{x^1=0} 
=
\f{\epsilon}{2r^2}[ r^4h_{AB}W^B_{,1}]_{,0}
	-\f{1}{2}\mc{D}_A (\ln r^2)_{,0} 
	}&&
\nonumber\\&&
	+\f{1}{2}(\mc{D}^E \ln r^2)h_{AE,0}
+\f{1}{2}\mc{D}^B h_{AB,0}
\end{eqnarray}
so that 
\begin{eqnarray}
\epsilon [  r^4h_{AB}W^B_{,1}]_{,0}
 &=&
 r^2\mc{D}_A (\ln r^2)_{,0} 
- {\mc{D}^B(r^2 h_{AB,0})}
\nonumber\\
&& {+2r^2R_{0A}|_{x^1=0} }
\end{eqnarray}
where the expression is evaluated at $x^1=0$.
\subsubsection{$R_{11}$}
Using \eqref{eq:conf_gAB} and \eqref{trgAB,a_quad} in \eqref{eq:R11_gen} gives
\begin{eqnarray}
R_{11}  
& = & -(\ln r^2 )_{,11} 
	+(\ln|g_{01}|)_{,1}(\ln r^2 )_{,1}  
	- \f{1}{2} [(\ln r^2)_{,1}]^2
\nonumber\\&&
	- \f{1}{4} h_{CB,1}h_{DA,1}h^{AC}h^{BD}
\end{eqnarray}
so that 
\begin{eqnarray}
0 & = &   r _{,11}
	-(r_{,1})   (\ln|g_{01}|)_{,1} 
\nonumber\\&&
	+ \f{r}{8} h_{CB,1}h_{DA,1}h^{AC}h^{BD} +\f{1}{2}rR_{11}\;\;.
	\nonumber\\
	\label{eq:R11_h}
\end{eqnarray}
So that for an affine null metric with $|g_{01}|=1$, we then have
\begin{eqnarray}
0 & = &   r _{,11}
	+ \f{r}{8} h_{CB,1}h_{DA,1}h^{AC}h^{BD} +\f{1}{2}rR_{11}
\end{eqnarray}
\subsubsection{$R_{1A}$}
Using  \eqref{eq:conf_gAB} and \eqref{DBg_AB,a} in \eqref{eq:R1A_gen} gives
 \begin{eqnarray}
R_{1A}  &=&
	-\f{\epsilon}{2r^2}(r^4 |g^{10}| h_{AB}U^B_{,1})_{,1}
	-\f{r^2 }{2} \Big[\f{ \mc{D}_A\ln|g_{10}|}{r^2}\Big]_{,1}
\nonumber\\&&
	+ {\f{1}{2}\mc{D}_A (\ln r^2)_{,1} 
	+	\f{1}{2r^2} \mc{D}^B (r^2 h_{AB,1}) }
\nonumber\\&&	
	- \mc{D}_A	(\ln r^2)_{,1}	
\end{eqnarray}
so that 
\begin{eqnarray}
R_{1A}  &=&
	-\f{\epsilon}{2r^2}(r^4 |g^{10}| h_{AB}U^B_{,1})_{,1}
	-\f{r^2 }{2} \Big[\f{ \mc{D}_A\ln|g_{10}|}{r^2}\Big]_{,1}
\nonumber\\&&
	-\mc{D}_A	(\ln r)_{,1}	
	+ (h_{AB,1}) \mc{D}^B(\ln  r )
	\nonumber\\&&
	+\f{1}{2}	  \mc{D}^B h_{AB,1} 
	\label{eq:R1A_h}
\end{eqnarray}
and with the notation of  \eqref{eq:spec_anm}, we have 
\begin{eqnarray}
R_{1A}  &=&
	-\f{\epsilon}{2r^2}(r^4  h_{AB}W^B_{,1})_{,1}
	-\mc{D}_A	(\ln r)_{,1}	
\nonumber\\&&
	+ (h_{AB,1}) \mc{D}^B(\ln  r )
	+\f{1}{2}	  \mc{D}^B h_{AB,1} 
\end{eqnarray}
\begin{widetext}
\subsubsection{$g_{01}g^{AB}R_{AB}$}
Using  \eqref{eq:conf_gAB}  in \eqref{eq:trRAB_gen} gives 
\begin{eqnarray}
0&=&
-g_{01}\sqrt{h} h^{AB}R^{(2)}_{AB}(r^2h_{AB})
+2\epsilon\sqrt{|g_{01}|}\sqrt{h}h^{AB} D_BD_A \sqrt{|g_{01}|}
	+ \Big[ \f{(r^4h )_{,0}}{r^2\sqrt{h}}\Big]_{,1} 
 	-\f{1}{2}\Big[\f{V(r^4 h)_{,1}}{r^2 \sqrt{h}}\Big]_{,1}
\nonumber\\&&
	 {+r^2\sqrt{h}D_C\Big[r^{-4}h^{-1} ( r^4h U^{C})_{,1}\Big]}
	+\f{1}{2}(g^{01})r^2\sqrt{h}r^2 h_{EF}	U^E_{,1}U^F_{,1}	
	+ g_{01}r^2\sqrt{h} g^{AB}R_{AB}
\end{eqnarray}
Since $h_{,0}=h_{,1}=D_C\sqrt{h}=0$ as well as \eqref{eq:conf_RS}, we have after factoring out $\sqrt{h}$ (which may also be zero at some coordinate points  -  and that choice depends only depends on the choise of `angular' coordinates)
\begin{eqnarray}
0&=&
-g_{01}  \mc{R}
+g_{01} \mc{D}^F\mc{D}_F \ln r^2 
+2\epsilon\sqrt{|g_{01}|} h^{AB} D_BD_A \sqrt{|g_{01}|}
	+ \Big[ \f{(r^4 )_{,0}}{r^2}\Big]_{,1} 
 	-\f{1}{2}\Big[\f{V(r^4 )_{,1}}{r^2 }\Big]_{,1}
	+r^2D_C\Big[\f{1}{r^{4}} ( r^4 U^{C})_{,1}\Big]
\nonumber\\&&
	+\f{1}{2}(g^{01})r^4 h_{EF}	U^E_{,1}U^F_{,1}	
	+ g_{01}h^{AB}R_{AB}
\end{eqnarray}
Simplification, rewriting the covariant derivatives of the scalars, co-vectors and vectors, while using $g_{01} = \epsilon|g_{01}|$ and $g^{01} = \epsilon|g^{01}|$,
gives us
\begin{eqnarray}
0&=& 
-\epsilon|g_{01}|  \mc{R}
+\epsilon|g_{01}| \mc{D}^F\mc{D}_F \ln r^2 
+2\epsilon\sqrt{|g_{01}|}\mc{D}^A\mc{D}_A \sqrt{|g_{01}|}
	+\Big[2  (r^2 )_{,0}
 	-V(r^2 )_{,1}\Big]_{,1}
	+\mc{D}_C\Big[\f{1}{r^{2}} ( r^4 U^{C})_{,1}\Big]
\nonumber\\&&
	+\f{\epsilon}{2}|g^{01}|r^4 h_{EF}	U^E_{,1}U^F_{,1}	
	+\epsilon |g_{01}| h^{AB}R_{AB} \label{eq:trRAB_h}
\end{eqnarray}
With the specifications to the affine null metric \eqref{eq:spec_anm}, there is after multiplication with $\epsilon$ ($\epsilon^2=1$)
\begin{eqnarray}
0&=& 
- \mc{R}
+ \mc{D}^F\mc{D}_F \ln r^2 
	+\Big[2 \epsilon (r^2 )_{,0}
 	+ W(r^2 )_{,1}\Big]_{,1}
	+\epsilon \mc{D}_C\Big[\f{( r^4 W^{C})_{,1}}{r^{2}} \Big]
	+\f{1}{2}r^4 h_{EF}	W^E_{,1}W^F_{,1}	
	+  h^{AB}R_{AB}\;\;.
\end{eqnarray}

\subsubsection{$g_{01}t^At^B R_{AB}$}
To find the respective 'conformal'  version of $t^At^B R_{AB}$ in \eqref{eq:ttRAB_gen}, consider first
\begin{eqnarray}
t^At^Bg_{AB,0}
&=&
	 r^2h_{AB,0} t^At^B \\
t^At^B g_{AB,1} & = &
	r^2h_{AB,1} t^At^B \\
t^At^B g_{AB,01} 
	&=&[(r^2)_{,1} h_{AB,0}+(r^2)_{,0} h_{AB,1} + r^2h_{AB,01}]t^At^B \\
(\ln \sqrt f)_{,0} &=& (\ln r^2)_{,0}  = \f{(r^2)_{,0}}{r^2}\\
t^At^B (\ln \sqrt f)_{,0}g_{AB,1} 
&=&
	[(r^2)_{,0}h_{AB,1} ]t^At^B \\
t^At^B (\ln \sqrt f)_{,1}g_{AB,1} &=& 	[(r^2)_{,1}h_{AB,1} ]t^At^B
\end{eqnarray}
which implies
\begin{eqnarray}
t^At^B \Big\{
	g_{AB,01}  
	-\f{1}{2}\Big[ (\ln\sqrt{f})_{,0} g_{AB,1}
	 + (\ln \sqrt{f})_{,1} g_{AB,0}\Big]
	 \Big\}
&=&
	t^At^B \Big[ r(rh_{AB})_{,01} 
	\Big]  
\end{eqnarray}
so that after insertion into \eqref{eq:ttRAB_gen}
\begin{eqnarray}
0&=&
  t^At^B\Bigg\{
	2\epsilon \sqrt{|g_{01}|}D_AD_B \sqrt{|g_{01}|}
	 +r(rh_{AB})_{,01}
 	 -\f{\sqrt{f}}{2}\Big[\f{Vg_{AB,1}}{\sqrt{f}}\Big]_{,1} 
	 +U^{C}D_Cg_{AB,1}
	+\f{1}{2}(D_C U^{C})g_{AB,1}
\nonumber\\&&
	+(\ln\sqrt{f})_{,1} D_{A}U_{B} 
	+g_{EA} D_{B}  U^E_{,1}
		 - g_{FA}g_{BC,1} \Big(D^CU^F - D^FU^C \Big)
+\f{1}{2}(g^{01})\Big[ U^E_{,1}U^F_{,1}g_{EA}g_{BF}
			\Big] 
			+ g_{01}R_{AB} 
	\Bigg\}\;\;.
	\nonumber\\
\end{eqnarray}
Next, inserting
\begin{eqnarray}
\f{\sqrt{f}t^At^B}{2}\Big[\f{Vg_{AB,1}}{\sqrt{f}}\Big]_{,1}  
	 & = &
 	\f{ t^At^B}{2}\Big[r^2Vh_{AB,1}\Big]_{,1} 
\end{eqnarray}
while using use $\sqrt{f} = r^2\sqrt{h}$  and modifying only  terms containing $g_{AB}$ gives
\begin{eqnarray}
 0&=&
  t^At^B\Bigg\{
	2\epsilon \sqrt{|g_{01}|}D_AD_B \sqrt{|g_{01}|}
	 +r(rh_{AB})_{,01}
 	 -\f{1}{2}\Big[r^2Vh_{AB,1}\Big]_{,1} 
	 +U^{C}D_Cg_{AB,1}
	+\f{1}{2}(D_C U^{C})g_{AB,1}
\nonumber\\&&
	+(\ln r^2)_{,1} D_{A}U_{B} 
	+r^2h_{EA} D_{B}  U^E_{,1}
		 - r^2h_{FA}g_{BC,1} \Big(D^CU^F - D^FU^C \Big)
	+\f{r^4}{2}(g^{01})h_{EA}h_{BF}  U^E_{,1}U^F_{,1}
\nonumber\\&&
		+ g_{01}R_{AB} 
	\Bigg\}\;\;.\label{eq:ev_tmp}
\end{eqnarray}
Inserting the relations 
\begin{eqnarray}
	2t^At^B\epsilon \sqrt{|g_{01}|}D_AD_B \sqrt{|g_{01}|}
&=&
2t^At^B\Big[r\epsilon \sqrt{|g_{01}|}\mc{D}_A\Big(\f{\mc{D}_B \sqrt{|g_{01}|}}{r}\Big)\Big]\\
\f{1}{2}(D_C U^{C})g_{AB,1}t^At^B
 & = & 
	  \f{1}{2}\Big[\mc{D}_C(r^2 U^{C})\Big] h_{AB,1} t^At^B \\
t^At^BU^{C}D_Cg_{AB,1}
	 & = & 
	t^At^BU^{C}\Big[ r^2\mc{D}_Ch_{AB,1}
	-h_{CA,1}\mc{D}_{B} r^2 
	+    h_{CA}h_{BH,1}\mc{D}^H  r^2\big]  
\\
 t^At^B(\ln r^2)_{,1} D_{A}U_{B}  
 & = &  
	 t^At^B( r^2)_{,1}  h_{BC} \mc{D}_A U^C  
\\
r^2t^At^Bh_{EA} D_{B}  U^E_{,1}
	 & = & 
	r^2t^At^B h_{CB}\mc{D}_A U^C_{,1} 
\\
- r^2h_{FA}g_{BC,1} \Big(D^CU^F - D^FU^C \Big)
&=&
 	- t^At^Br^2h_{FA}h_{BC,1} \Big(\mc{D}^C U^F -  \mc{D}^F U^C   \Big)   
\end{eqnarray}
into \eqref{eq:ev_tmp} yields
\begin{eqnarray}
0&=&  t^At^B\Bigg\{
	2r\epsilon \sqrt{|g_{01}|}\mc{D}_A\Big(\f{\mc{D}_B \sqrt{|g_{01}|}}{r}\Big)
	 +r(rh_{AB})_{,01}
 	 -\f{1}{2}\Big[r^2Vh_{AB,1}\Big]_{,1} 
	 +r^2U^{C}\mc{D}_Ch_{AB,1}
	-U^{C}h_{CA,1}\mc{D}_{B} r^2 
\nonumber\\&&
	\qquad\;\;	
	+    U^{C}h_{CA}h_{BH,1}\mc{D}^H  r^2
	+\f{1}{2}\Big[\mc{D}_C(r^2 U^{C})\Big] h_{AB,1}
	+( r^2)_{,1}  h_{BC} \mc{D}_A U^C  
	+r^2h_{EA} \mc{D}_{B}  U^E_{,1}
\nonumber\\&&
	\qquad\;\;
	-r^2h_{FA}h_{BC,1} \Big(\mc{D}^C U^F -  \mc{D}^F U^C   \Big)   
	+\f{r^4}{2}(g^{01})h_{EA}h_{BF}  U^E_{,1}U^F_{,1}
		+ g_{01}R_{AB} 
	\Bigg\}
\end{eqnarray}
Further simplification, while using  $g^{01} = \epsilon|g^{01}|$, and $t^A = m^A/r$ results in
\begin{eqnarray}
0&=&
 m^Am^B\Bigg\{
	2r\epsilon \sqrt{|g_{01}|}\mc{D}_A\Big(\f{\mc{D}_B \sqrt{|g_{01}|}}{r}\Big)
	 +r(rh_{AB})_{,01}
 	 -\f{1}{2}\Big[r^2Vh_{AB,1}\Big]_{,1} 
	 +r^2U^{C}\mc{D}_Ch_{AB,1}
	+\f{1}{2}\Big[\mc{D}_C(r^2 U^{C})\Big] h_{AB,1}
\nonumber\\&&
\qquad	+( r^2)_{,1}  h_{BC} \mc{D}_A U^C  
	+r^2h_{EA} \mc{D}_{B}  U^E_{,1}
	-r^2h_{FA}h_{BC,1} \Big(\mc{D}^C U^F -  \mc{D}^F U^C   \Big)   
	+\f{\epsilon r^4}{2}|g^{01}|h_{EA}h_{BF}  U^E_{,1}U^F_{,1}
\nonumber\\&&
	\qquad\;\;
 		+\epsilon| g_{01}|R_{AB} \;\;.
	\Bigg\}\label{eq:ttRAB_h}
\end{eqnarray}
With the specifications for the affine, null metric \eqref{eq:spec_anm},  we have 
\begin{eqnarray}
0&=&
 m^Am^B\Bigg\{
	 r(rh_{AB})_{,01}
 	 +\f{\epsilon}{2}\Big[r^2 W h_{AB,1}\Big]_{,1} 
	 +r^2W^{C}\mc{D}_Ch_{AB,1}
	+\f{1}{2}\Big[\mc{D}_C(r^2 W^{C})\Big] h_{AB,1}
	+( r^2)_{,1}  h_{BC} \mc{D}_A W^C  
\nonumber\\&&
\qquad	
	+r^2h_{EA} \mc{D}_{B}  W^E_{,1}
	-r^2h_{FA}h_{BC,1} \Big(\mc{D}^C W^F -  \mc{D}^F W^C   \Big)   
	+\f{\epsilon r^4}{2} h_{EA}h_{BF}  W^E_{,1}W^F_{,1}
 		+\epsilon R_{AB} 
	\Bigg\}
\end{eqnarray}

\end{widetext}

\end{appendix}

\end{document}